# Sustainability assessment of 4G and 5G universal mobile broadband strategies

Edward J. Oughton[1,2,*], Jeongjin Oh[3], Sara Ballan[3], Julius Kusuma[4]

[1]College of Science, George Mason University, Fairfax, VA, USA

[2]Environmental Change Institute, University of Oxford, Oxford, Oxfordshire, UK

[3]World Bank, Washington DC, USA

[4]Meta, Menlo Park, USA

*Corresponding author: Edward J. Oughton (e-mail: eoughton@gmu.edu)

Address: College of Science, George Mason University, 4400 University Drive, Fairfax, VA

## Abstract

With infrastructure systems lasting for decades, even centuries, there is growing need to assess sustainability impacts. However, compared to energy or transportation networks (which each contribute roughly one third of global emissions), broadband networks have arguably received less attention due to their much smaller footprint (~1.8-3.9% of global emissions). Nevertheless, many countries are looking to provide universal mobile broadband over the next decade to meet Goal 9 of the Sustainable Development Goals (SDGs). Therefore, this paper evaluates the future sustainability impacts of providing either 4G or 5G mobile broadband, quantifying the carbon and other environmental emissions associated with each universal broadband strategy. This paper contributes the first *ex ante* sustainability assessment of global universal mobile broadband strategies aimed at delivering SDG Goal 9.

# 1. Introduction

Broadband-enabled technologies, such as smartphones, play a critical role in the climate agenda, with both positive and negative effects on greenhouse gas (GHG) emissions (Haldar and Sethi, 2022). For example, smartphones enable a wide range of services and applications that can help to reduce a citizen's environmental impact. However, expanding the availability of broadband infrastructure can have a direct negative effect on the environment, as broadband assets require electricity to function, with the production of this energy from non-renewable sources releasing carbon and other atmospheric emissions into the environment (Farquharson, Jaramillo and Samaras, 2018).

A meta-analysis of the literature suggests that globally the use of Information Communication Technologies (ICTs) accounts for 1.8%, up to as much as 3.9%, of global GHG emissions, with these ranges produced by heterogenous assessment assumptions, and system boundary definitions (Freitag et al., 2021). Depending on the method, emissions are roughly split equally three ways between networks, data centers and user devices (Andrae and Edler, 2015; Belkhir and Elmeligi, 2018; Malmodin and Lundén, 2018). Indeed, with our dependence on these technologies rapidly growing, there are concerns that increasing computational requirements may lead to greater energy consumption (Jones, 2018; Masanet et al., 2020; Gupta et al., 2022). This highlights the need for greater consideration of different mitigation impacts, to prevent unsustainable paths being locked in over future decades (Kaack et al., 2022).

Current estimates put the number of global Internet users at approximately 66%, representing 5.3 billion people, having increased from only 54% in 2019 (International Telecommunication Union and United Nations Educational, Scientific and Cultural Organization, 2022). Much of this increase (approximately 11%) was due to rapid adoption in Low- and Middle-Income Countries (LMICs) during the coronavirus pandemic, as new users needed to access a range of online services to learn, work, access remote healthcare and government services (Mac Domhnaill, Mohan and McCoy, 2021; Whitacre, 2021; Zhang, 2021; Isley and Low, 2022; Abrardi and Sabatino, 2023). However, this still leaves approximately 2.7 billion people unconnected. Goal 9 of the Sustainable Development Goals (SDGs) aims to provide affordable universal infrastructure for inclusive and sustainable industrialization (United Nations, 2023). Mobile broadband infrastructure is central to this goal, with the motivation being that there are a range of economic, social and environmental benefits from citizens gaining Internet connectivity (Hasbi, 2020; Ndubuisi, Otioma and Tetteh, 2021; Deller, Whitacre and Conroy, 2022; Stephens, Mack and Mann, 2022; Chen *et al.*, 2023).

Additionally, in many LMICs, access to mobile broadband has been expedited by the expansion of cellular sites into areas either not connected to a national grid, or connected but receiving unreliable electrical power. Many of these rural sites may be considered "off-grid" or "bad-grid", which tend to utilize on-site diesel-powered generators for their operation, inflicting a larger environmental impact than other energy sources. Moreover, the building of future broadband infrastructure will increase the level of energy consumption taking place from the overall quantity of network assets. Therefore, it is imperative that this dynamic is modeled and understood, to provide strategic insight for different global mobile universal broadband strategies.

Consequently, four key research questions will be investigated in this analysis, which include:

1. What are the financial costs of universal mobile broadband strategies, for different technologies, business models and policy options?
2. What are the energy and environmental impacts of universal mobile broadband strategies?
3. How do different business model options affect the energy and environmental impacts of universal mobile broadband strategies?
4. To what extent do environmental benefits arise from shifting off-grid diesel sites to renewable power?

In order to answer this set of research questions, the structure of this paper will be as follows. First, a literature review of relevant research will be undertaken in Section 2, followed by the articulation of a method capable of answering the stipulated research questions in Section 3. Next, the results will be presented in Section 4 and then discussed in Section 5. Finally, conclusions will be provided in Section 6, highlighting the key literature contribution and areas of future research.

## 2. Literature review

Here relevant studies pertaining to the research questions will be appraised. For example, a recent review of the literature identifies that there is a significant lack of peer-reviewed, transparent, up-to-date, and open-access network assessments for operational energy efficiency in cellular networks, such as for 4G or 5G (Mytton and Ashtine, 2022; Williams, Sovacool and Foxon, 2022). Indeed, one systematic review of over 100 papers, examining the environmental impacts of ICT between 1978 and 2022, finds that many have incompatible methodologies, preventing sensible comparisons being drawn from the range of available estimates (Zhang and Wei, 2022). This provides important motivation for the subject of this paper. In this particular literature review, the content is evaluated for each of the research question areas, which include (i) investment magnitude, (ii) business models, and (iii) energy and environmental impacts.

### 2.1. Investment estimates for mobile universal broadband

To achieve SDG 9 and deliver universal mobile broadband, a substantial quantity of global capital would need to be deployed, with the associated magnitude varying depending on a range of existing estimates. The challenge is that many methods may vary in their chosen assumptions and model approaches, making it hard to readily compare results. One of the most notable examples is from the Connecting Humanity report, which examines the investment needs for providing broadband service to currently unconnected users globally, by estimating aggregate costs (International Telecommunication Union, 2020). Universal broadband is defined in the study as delivering the equivalent of a 4G mobile connection (~10 Mbps in speed) to all those over 10 years of age. An estimated total cost of $428 billion is calculated to connect all unconnected users by 2030, with this figure comprised of approximately $104 billion on mobile capital expenditure (capex), $70 billion on metro and backbone fiber, $140 billion of mobile operational expenditure (opex), $70 billion on remote satellite coverage, $6 billion on policy and regulatory costs, and $40 billion on ICT skills and content. This investment is broken down by $135 billion in South Asia, $97 billion in Sub-Saharan Africa (SSA), $83 billion in East Asia & Pacific, $51 billion in the Americas, $33 billion in Europe and Central Asia, and $28 billion in the Middle East and North Africa (MENA). Low Income Countries (LICs) are estimated to require 17% of the required capital, versus 52% in Lower Middle-Income Countries (LMCs), 25% in Upper Middle-Income Countries (UMCs), and 6% in High Income Countries (HICs).

However, one study by the International Monetary Fund replicates this analysis using a spatially disaggregated model accounting for each country's demographic growth trend, heterogenous population density characteristics, and future economic characteristics (Oughton, Amaglobeli and Moszoro, 2023). The total estimated investment for universal broadband provision for all currently unconnected users under the baseline is estimated to be $418 billion to connect unconnected users (0.45% of global GDP). The analysis emphasizes how changes to basic model inputs can vastly change the estimated investment, highlighting the need for future broadband policy assessments to be very explicit about the quantity of data each user can consume, as well as other Quality of Service (QoS) parameters. A key limitation of both of these reported studies is that they exclude the investment required to upgrade underserved areas, such as those with a very basic service (and thus are connected, but do not provide sufficient service to enable satisfactory usage).

Indeed, a separate analysis, focusing on evaluating universal 4G and 5G infrastructure strategies in LMICs, uses a very different set of methodological assumptions (Oughton *et al.*, 2022). Instead of focusing on only unconnected users, the model utilized also considers those who may be *underserved* by their existing connection. For a range of highly ambitious headline speed targets (up to ~25 Mbps, up to ~200 Mbps, and up to ~400 Mbps per user), the estimated investment is as high as $2 trillion to deploy and operate 5G non-standalone over the next decade. The key difference is that this appraisal does not predominantly isolate itself to areas of market failure, meaning this estimate represents both public and private investment over the next decade. Finally, while this value may be very large, modeling results suggest that this amount could be almost halved if a encouraging regulatory environment is created, favorable to infrastructure investment, with fiscal and regulatory regimes conducive to lowering cost. Indeed, infrastructure sharing strategies, along with changes to spectrum license costs and corporate taxation rates, were central to these cost reductions.

## 2.2. Infrastructure sharing business models

A variety of infrastructure sharing business models are available to operators and governments, with the dual ability to reduce costs and emissions, by avoiding infrastructure duplication.

Although there are a range of different sharing options, four strategies are identified here based on the available literature (Koratagere Anantha Kumar and Oughton, 2023; Kumar and Oughton, 2023). Firstly, a business-as-usual case is where each operator builds and manages their own network, which is the status quo in many markets around the world. However, with static subscription growth and rising costs in many countries, numerous operators and governments have been examining passive infrastructure sharing as a way to reduce costs, while maintaining a degree of Radio Access Network (RAN) separation. This strategy would involve operators sharing a site compound and the civil engineering tower structure, to mount their own separate radio electronics. Where greater costs savings may be desired, operators can explore active sharing where at least two competitors share both passive and active radio electronic assets. For example, this may also include the sharing of antennas, radios, and Baseband Units (BBU). Such an approach is often called a Multi-Operator Radio Access Network (MORAN) and Multi-Operator Core Network (MOCN), depending on the level of sharing. One more basic form of active sharing is simple roam agreements between competitors, to allow users to gain extend service coverage. Finally, other more heterogenous sharing strategies include examples such as a Shared Rural Network (SRN). In this circumstance, competitors would maintain their own networks in urban and suburban areas (preserving infrastructure competition in high demand locations), while sharing infrastructure in less viable rural and remote regions (where separate network deployment may not be so viable).

One continental assessment of Africa focused on the cost of serving unconnected and unserved users using 4G mobile broadband (Oughton, 2023). The initial estimate is approximately $0.4 trillion to deliver 10 GB/Month per user, with this quantity rising to $0.7-1 trillion should a higher traffic target be adopted for 30 GB/Month. Interestingly, infrastructure sharing business models were assessed, with active infrastructure sharing delivering impressive savings (48–67 % for 10 GB/Month and 64–78 % for 30 GB/Month), compared to more modest savings when considering only non-electronic passive components (18–34 % for 10 GB/Month and of 26–44 % for 30 GB/Month). However, when deploying a Shared rural network to balance out infrastructure competition in urban and suburban areas, and increased sharing in less viable rural and remote areas, up to 52 % of the cost saving could be achieved.

Now the environmental impacts of broadband infrastructure will be discussed.

## 2.3. Energy and environmental impacts of universal broadband

An assessment of European telecom network operators finds that for 2015-2018 annual electricity consumption remained nearly constant (Lundén *et al.*, 2022). While total subscribers increased by 3%, only a 1% increase took place in electricity consumption, despite traffic increasing by a factor of three. Moreover, when considering a longer time period in the same study, per subscription electricity consumption remained static (~30 kWh/subscription), despite data traffic increasing by a factor of twelve. Therefore, what we can conclude from this assessment, is that existing broadband networks can be readily increased in traffic capacity without a commensurate increase in energy consumption. Indeed, energy consumption increases are therefore driven by increasing the coverage of an existing network, as this involves building new assets.

The increasing use of digital technologies is found to have mixed effects on energy consumption. Indeed, there are four key energy-economic impact effects which arise, including (1) direct effects from the physical production, usage and disposal of digital devices and equipment, (2) energy efficiency gains from increasing use of digital technologies, (3) consequential improvements in economic growth and productivity resulting from labor and capital efficiencies and (4) macroeconomic structural changes resulting from new ICT services. The first and third effects are found to lead to an increase in energy consumption, whereas the second and forth effects are found to lead to a decrease in energy consumption (Lange, Pohl and Santarius, 2020). Indeed, the authors of this study suggest that an overall increase in energy consumption takes place, because the two increasing effects are found to prevail based on empirical and theoretical findings. For example, the total energy consumption of the ICT sector has been estimated at approximately 805 TWh (~3.6%), comprised of 304 TWh from user devices, 220 TWh from ICT networks, and 245 TWh from data centers (based on global electricity consumption of 21,000 TWh) (Malmodin and Lundén, 2018). These estimates equate to carbon emissions of 730 Mt (1.4%), comprised of 395 Mt from user devices, 180 Mt from ICT networks, and 180 Mt from data centers (based a global carbon footprint of 53 Gt $CO_2$).

Importantly, while there have been a significant number of studies focusing on user devices (Clément, Jacquemotte and Hilty, 2020), or data centers (Siddik, Shehabi and Marston, 2021), there is a comparative lack of whole-system assessments of networks, highlighting the need for the research presented in this paper.

One country-specific assessment evaluates the environmental impact of 4G LTE networks in Peru, focusing on the sustainability implications of connecting underserved regions. The findings estimate total annual carbon emissions to be approximately 81-103 kg $CO_2$ eq. per subscription, which is broadly equivalent to 1.35–1.73 kg $CO_2$ eq./GB. However, between 68-86% of these estimates arise from embodied emissions in end-user devices, whereas operational emissions were only approximately one-third of the total, primarily due to end user devices consuming electricity, and to a lower extent by access networks and data centers. Generally, a linear correlation is found between operational carbon emissions and the number of subscribers (Ruiz *et al.*, 2022). Yet, networks serving very sparsely populated locations have higher energy consumption and carbon emissions on a per subscriber basis when compared to urban users.

With coverage in rural and remote areas often affected by access to electricity distribution networks, many sites globally have been installed utilizing diesel-powered generators to provide electricity (Tweed, 2013). Unfortunately, such an approach which intensely utilizes fossil-based fuels does contribute to air quality issues arising from the production of emissions including carbon dioxide ($CO_2$), particulate matter (PM2.5), sulfur oxides (SOx), nitrogen oxides (NOx) (Farquharson, Jaramillo and Samaras, 2018). Therefore, increasingly renewable energy sources are being explored to provide power to cell sites with greatly reduced pollution (and ideally at lower cost when compared to driving diesel fuel to each site on a weekly basis) (Suman and De, 2020). For example, one study evaluated the cost of utilizing diesel generation against solar-enabled power systems, finding that it is possible to transition (Ahmad et al., 2015). There is a significant cost added to operations from diesel transportation, meaning the cost of powering cell sites this way increases with distance from urban centers. Thus, making renewable power options more attractive for rural and remote areas.

In another assessment, load balancing optimization is demonstrated to achieve a zero-outage probability for cell sites when considering a variety of renewable energy approaches, including solar photovoltaics, wind turbines, and biomass generators (Hossain et al., 2020). Indeed, such an approach can achieve a per-unit cost of energy of $0.38 kWh, with the potential to decrease emissions to as low as 0.38 kg per year (Hossain et al., 2021). Detailed site assessment has also indicated that utilizing renewable energy sources in rural areas is an effective emissions-reduction strategy for transitioning from conventional diesel-powered energy generation at cell sites, while managing to minimize total cost (Amole *et al.*, 2021; Baidas *et al.*, 2022; Islam *et al.*, 2023). The use of software defined network techniques is identified as a key approach to help balance capacity and power demand (Jahid *et al.*, 2021), particularly as networks are upgraded to 5G over the next decade.

Having evaluated literature relevant to the stated research questions, the method will now be articulated.

## 3. Method

The approach adopted here utilizes a techno-economic scenario-based method to provide insight on 'what if' questions, pertaining to the environmental impacts of universal mobile broadband infrastructure. The use of scenarios is common in situations where incomplete scientific data are available to undertake rigorous and realistic modeling, yet where analytics are essential requirements to support decisions (Elzen *et al.*, 2002; Crawford, 2019; Frith and Tapinos, 2020; Gordon, 2020; Jefferson, 2020). For example, in the field of infrastructure analysis (Hall *et al.*, 2016, 2017; Thoung *et al.*, 2016; Oughton *et al.*, 2019; Pant *et al.*, 2020; Schneir *et al.*, 2023). Commonly when modeling is used to support mobile broadband decisions in telecommunication regulators, a hypothetical Mobile Network Operator (MNO) is developed with a representative set of characteristics (users, sites, spectrum, market share etc.), based on a Long-Run Incremental Cost approach (Ofcom, 2018).

The approach taken here is to divide each country's local statistical areas into deciles based on population density, to group together the key factor affecting the cost of delivery. This approach makes the global modeling task more tractable, reducing ~10,000 level 1 and 2 local statistical areas globally (GADM, 2019), down to approximately ~1,900.

The method adopted here is illustrated in box model form in Figure 1.

Figure 1 A box diagram illustration of the sequential modeling method

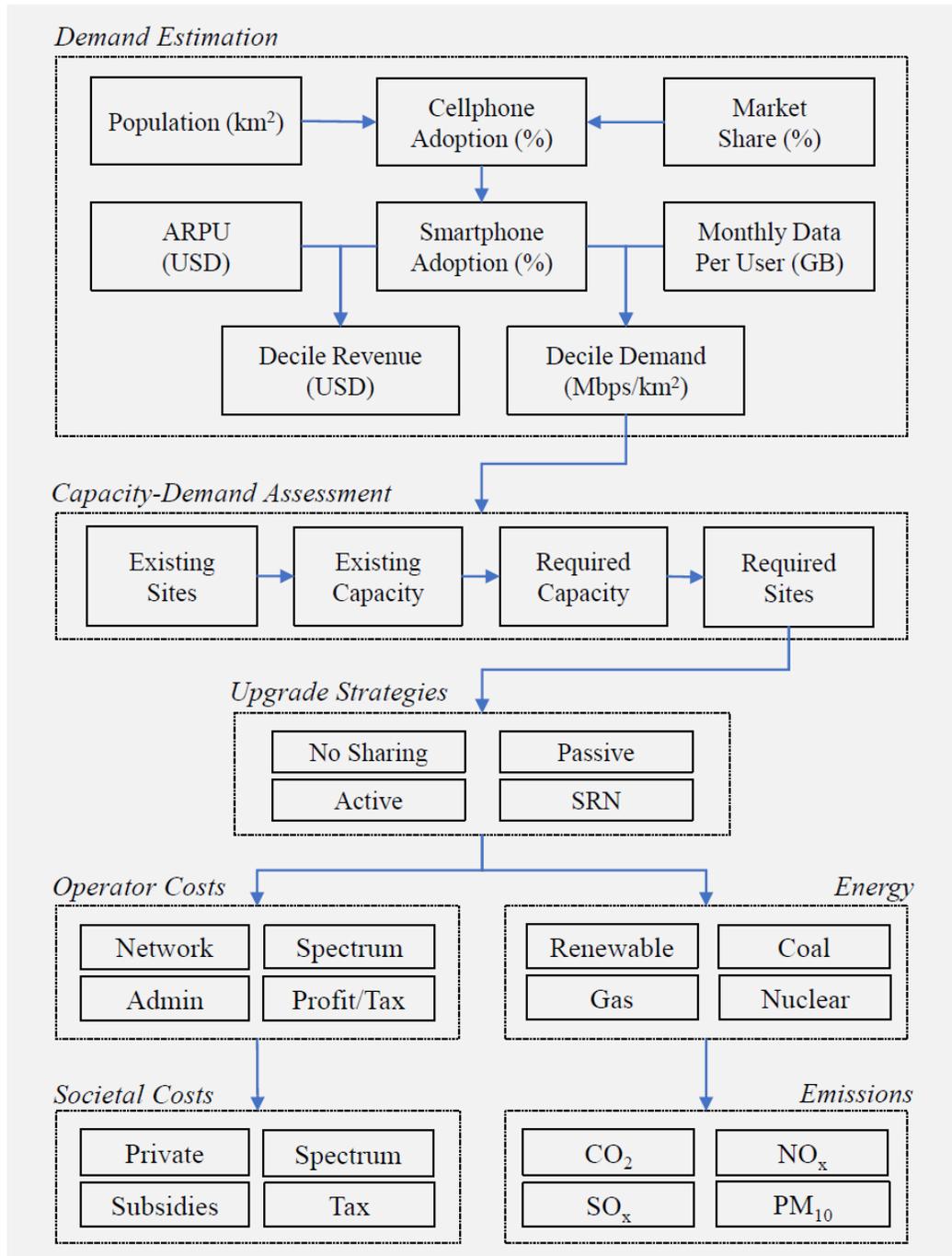

## 3.1. Strategies

Four different **technology strategies** are adopted. These involve using either 4G or 5G to provide wide-area mobile broadband services. Each site requires a connection from the cell site to a fiber Point of Presence (PoP), treated here as being either a fixed fiber connection (F) or microwave wireless link (W). While a fiber backhaul provides higher-capacity and greater reliability, this option is likely to be much more expensive than using a wireless backhaul connection, especially for rural and remote areas (Oughton, Boch and Kusuma, 2022). A key caveat is that for 5G services to be provided and utilized, users would need to have a 5G device to connect to the network (which are generally more expensive than older 4G devices).

In terms of business models, four different **infrastructure sharing strategies** are selected. The business-as-usual baseline is focused on the standard approach to building mobile telecommunication infrastructure, where an MNO will build and operator their own exclusive network assets to maximize their own exclusive coverage area (and thus, potential market share). In the passive infrastructure sharing strategy, this is where multiple MNOs share site compounds and civil infrastructure (the passive non-electronic assets), but still deploy their own active infrastructure components (radio heads, baseband units and other processing equipment). Moreover, in the active infrastructure sharing strategy, multiple MNOs share both passive assets (site compounds, towers etc.), as well as active electronic radio equipment (antennas, baseband units and other processing equipment). Finally, in the SRN strategy MNOs implement the business-as-usual baseline approach of building their own exclusive infrastructure in urban and suburban locations, whereas they actively share non-electronic and electronic assets in rural areas.

In terms of **policy strategies**, five key options are explored. Firstly, the baseline strategy consists of using data-driven values for both taxation and spectrum fees. In addition to strategies which vary these values into low and high variants, to explore the impacts on overall financial cost from fluctuating these model input parameters.

Finally, in addition to quantifying emissions impacts for the baseline and infrastructure sharing approaches, a single **emissions reduction strategy** will also be utilized to explore the impact of off-grid power sources, and converting diesel-powered generators to renewable power generation in hard-to-reach locations.

## 3.2. Scenarios

Three different **capacity target scenarios** are utilized to capture future data traffic changes. Indeed, costs are highly correlated with data consumption, as more data traffic will require additional sites to be built which require investment to construct and operate (due to energy consumption, maintenance etc.). Therefore, a baseline of 30 GB/Month is used for the future year of 2030, commensurate with existing mobile traffic forecasts (Ericsson, 2023). In contrast, the high scenario is set to 40 GB/Month (representing a more optimistic usage outcome), and the low to 20 GB/Month (representing a more pessimistic usage outcome), to capture variability in data traffic quantities.

Three different **adoption scenarios** are specific, developed based on future potential increases in adoption by country income group. For example, in HICs where adoption is already mature and growth slow, 0.5%, 1% and 1.5% rates are used for the low, baseline and high scenarios, respectively, with 1%, 2% and 4% in UMCs. This is compared to higher rates in LICs where adoption is still growing, for example, using 1.5%, 3% and 6% for LMCs, and 2%, 4% and 6% in LICs for the low, baseline and high adoption scenarios, respectively (equating to mean global values of 1.3%, 2.5% and 4.4% for the different scenarios).

## 3.3. General system model - Demand

Prior to obtaining the data traffic demand for a local statistical area, it is necessary to convert from the capacity target scenario value ($S_t$) in Gigabytes (GB) per month ($t$) per user, into a per second value. For example, the data rate per user in Megabits Per Second (Mbps) ($Cap_i$) is obtained using Equation (1), firstly converting the monthly traffic quantity from GB to MB (multiplying by 1000), and then converting from bytes to bits (multiplying by 8), obtaining the traffic in Megabits per month. Next, the quantity of days per month ($n_d$) (30) can be used to convert from Megabits per month to Megabits per day. Then the quantity of Megabits per hour in the busy hour ($BH$) can be obtained based on 15% of traffic being exchanged in the busiest hour. Finally, the Megabits per hour value can be converted to the per second quantity in Megabits per second (Mbps) based on 3,600 seconds per hour.

$$Cap_i = S_t \cdot 1000 \cdot 8 \cdot n_d^{-1} \cdot BH^{-1} \cdot 3600^{-1} \qquad (1)$$

Now the data rate per user has been estimated ($Cap_i$), it is possible to convert this to an area traffic demand ($Demand_i$) (Mbps km²) in the $i$th local region at time $t$, as per Equation (2). Firstly, a set of unique local smartphone users can be estimated using the total population ($Pop_i$) in the area, a unique cell phone penetration value ($Pen_{it}$) and a unique smartphone penetration value by urban-rural area ($SP_{Pen_{it}}$). Next these smartphone users can be multiplied by data rate per user in the busiest hour of the day ($Cap_i$), before dividing based on the market share of the hypothetical network operators being modeled ($Market_{Share_i}$). Finally, the maximum traffic value is selected over the time horizon modeled (up to 2030), before converting to a traffic density (Mbps/km²) given the geographic area (km²) of the local statistical area being modeled.

$$Demand_i = Max((\frac{Pop_i \cdot Pen_{it} \cdot SP_{Pen_{it}} \cdot Cap_i}{Market_{Share_i}})/Area_i) \qquad (2)$$

Next, the generated revenue from each local statistical area is estimated ($Revenue_i$) utilizing an Average Revenue Per User ($ARPU_i$) value, substituted for the data rate per user, detailed as follows in Equation (3).

$$Revenue_i = \frac{Pop_i \cdot Pen_{it} \cdot SP_{Pen_{it}} \cdot ARPU_i}{Market\_Share_i} \qquad (3)$$

ARPU data values are adapted from GSMA data using country income groups (GSMA, 2020). For example, the low, baseline and high ARPU values are segmented for HICs ($10, $15, $20), UMICs ($8, $12, $16), LMICs ($6, $10, $14), and LICs ($3, $5, $10). High values are allocated to urban areas, baseline values to suburban and low values to rural areas. Market share data is specified based on the number of major MNOs present in crowdsourced OpenCelliD data for each country (OpenCelliD, 2022), for example, with four major operators the market share would be 25%. Using consumer price projection data (International Monetary Fund, 2021), the revenue over the assessment period is converted to the Present Value (PV) via a 5% discount rate to capture the time value of money.

## 3.4. General system model - Supply

The capacity of a wireless cellular network can be estimated based on the density of sites (per square kilometer), the generation of technology (e.g., 4G or 5G) and associated efficiency, and the available spectrum (e.g., 10-40 MHz channel bandwidth). Given these three key factors, it is possible to estimate each $i$th decile capacity ($C_{it}$) (Mbps/km²) in time $t$, by summing for all $f$ frequencies in use, given the product of the mean area spectral efficiency ($\eta_{Area}^{f}$) (bits/Hz) and channel bandwidth ($BW^f$) (Hz), as specified in Equation (4):

$$C_{it} = \sum_{f} \eta_{Area}^{f} BW^{f} \quad (4)$$

Importantly, the spectrum portfolio for each country informs the bandwidth of the available frequencies being modeled ($BW^f$), for a hypothetical mobile network operator (Frias, Mendo and Oughton, 2020). This is possible because data demand is lower in LICs, leading to smaller spectrum portfolios. And vice versa in HICs, where data demand is greater, and spectrum portfolios are larger.

The mean spectral efficiency ($\bar{\eta}_{Area}^{f}$) of an area is able to be modeled by treating radio wave propagation as a stochastic process. Indeed, this accurately represents real world networks where different types of environmental clutter affect the path of propagated waves (e.g., buildings, trees, other concrete and steel structures etc.). Therefore, the mean area spectral efficiency ($\bar{\eta}_{Area}^{f}$) can be estimated using the mean number of cells per site ($\bar{\eta}_{cells}^{f}$) (e.g., usually three for tri-sectored cell sites), and the cell sites density of transmitting antennas in the frequency being modeled ($\bar{\eta}_{sites}^{f}$). Equation (5) specifies the approach.

$$\bar{\eta}_{Area}^{f} = \bar{\eta}_{cells}^{f} \cdot \bar{\eta}_{sites}^{f} \quad (5)$$

The spectral efficiency of a single radio link can be estimated by accounting for the received signal at a piece of User Equipment (UE), such as a smartphone, along with the co-channel interference from nearby transmitters and UE noise, to obtain the Signal-to-Interference-and-Noise-Ratio (SINR).

For example, the received signal ($Signal_i$) for the $i$th link at the UE is estimated as detailed in Equation (6), which contains the transmitter power ($TxPower_i$) (dBm), transmitter gain ($TxGain_i$) (dB), any transmitter losses ($TxLosses_i$) (dB), the path loss ($PathLoss_i$) (dB), receiver antenna gain ($RxGain_i$) (dB), and finally any receiver losses ($RxLosses_i$) (dB).

$$Signal_i = TxPower_i + TxGain_i - TxLosses_i - PathLoss_i + RxGain_i - RxLosses_i \qquad (6)$$

Next, this same equation can be used to estimate the interference ($Interference_{i,j}$) for the $i$th path from the $j$th cell, at the UE from all nearby transmitting antennas. The associated noise ($Noise_i$) for the $i$th path can be estimated, as in Equation (7), given the Boltzmann constant ($k$), the temperature in Kelvin ($T$), the frequency bandwidth ($BW$) and the UE noise flood ($NF$).

$$Noise_i = 10log_{10}\,(k\,T\,1000) + NF + 10log_{10}(BW) \qquad (7)$$

Finally, the SINR for the $i$th path can be estimated ($SINR_i$), given the estimated received signal ($Signal_i$), interference ($Interference_{i,j}$) and noise ($Noise_i$) values.

$$SINR_i = \frac{Signal_i}{\sum_j (Interference_{i,j} + Noise_i)} \qquad (8)$$

This SINR value is then mapped to a spectral efficiency in a known lookup table (ETSI, 2018). Generally, the SINR is strongly affected by the spatial-aspects associated with distance to and from nearby cell site antennas (for both wanted received signal and unwanted interference). Commonly, the stochastic Monte Carlo approach is based on estimating probability density function performance curves for different cell site densities (Mogensen *et al.*, 2007; Frias, González-Valderrama and Pérez Martínez, 2017; Oughton and Jha, 2021). To model the path between the transmitter and receiver, the ITU report SM.2028-2 is followed using Monte Carlo simulation for modeling radio services (ITU-R, 2017). A free space path loss model is adopted with a log normal shadow fading distribution, with distances over 500 meters treated as non-Line of Sight (LoS). Very often the QoS a network operator will target will be in the 90% reliability level, which can be utilized here, and obtained from the probability density function performance curves. Here, the developed codebase is used to generate a set of site density to capacity lookup tables, enabling the capacity of a local statistical area to be estimated, along with the number of new sites required to meet future demand.

For example, if the existing capacity is below the required traffic demand, the number of sites needing to be built can be estimated ($NewSites_i$) for the $i$th local statistical area. To do so, the existing number of sites ($ExistingSites_i$) is subtracted from the total number of sites required to meet future traffic demand ($TotalSites_i$). As detailed in Equation (9):

$$NewSites_i = TotalSites_i - ExistingSites_i \qquad (9)$$

The cost of building these new assets is estimated using a cost model for the hypothetical mobile network operator being evaluated.

### 3.5. General system model - Costs

When the mobile cellular network needs to be expanded to provide more capacity, MNOs have the option to build entirely new greenfield sites, or they can upgrade any existing sites present within each decile. To do so when building a new site in the $i$th decile, the network cost ($Network_i$) can be calculated by summing the investment in the Radio Access Network (RAN) ($Equipment_i$), backhaul ($Backhaul_i$), civil engineering for the tower structure ($Civils_i$) and the core network ($Core_i$), as detailed in Equation 10. In contrast, when upgrading sites in a decile, an MNO has the same cost structure as Equation (10), except no further investment is required in building the civil engineering tower structure (as a tower structure already exists).

$$Network_i = Equipment_i + Backhaul_i + Civils_i + Core_i \qquad (10)$$

Next, the overall cost for a decile can be estimated, based on the required private investment to build and upgrade sites present, as well as any necessary government subsidies. Thus, the private cost ($PrivateCost_i$) an MNO will incur in the $i$th decile will be comprised of the network investment cost ($Network_i$), administration costs for operation ($Administration_i$), spectrum license fees ($Spectrum_i$), taxes ($Taxes_i$), and profits ($Profit_i$.), as per Equation (11).

$$PrivateCost_i = Network_i + Administration_i + Spectrum_i + Taxes_i + Profit_i. \qquad (11)$$

Whereas the net government cost ($GovernmentCost_i$) in the $i$th decile is treated as the required state subsidy ($Subsidy_i$) minus any revenues gained from spectrum fees ($Spectrum_i$) and taxation ($Tax_i$), as per Equation (12).

$$GovernmentCost_i = Subsidy_i - (Spectrum_i + Tax_i) \qquad (12)$$

Therefore, the final financial cost ($FinancialCost_i$) to society can be estimated in each decile by adding together the private cost and net government cost ($FinancialCost_i = PrivateCost_i + GovernmentCost_i$). Summing the cost for all deciles provided the total country cost.

## 3.6. General system model – Energy and emissions

To calculate the annual energy consumption ($Energy_i$) (kWh) of all assets in operation in the $i$th region, Equation (13) is utilized. This involves multiplying the number of existing ($ExistingSites_i$) and new sites ($NewSites_i$), by the average hourly energy consumption per site ($EnergyPerSite$), and the number of hours in one year:

$$Energy_i = (ExistingSites_i + NewSites_i) \cdot EnergyPerSite \cdot (24 \cdot 365) \tag{13}$$

The average hourly energy consumption per site is taken from the literature based on empirical field measurement campaigns, where a macro cell site utilizes approximately 0.249 kWh of electricity (Kusuma *et al.*, 2022). A traditional monolithic basestation has multiple transceivers consisting of a power amplifier, Radio Frequency (RF) small-signal transceiver module, and a baseband engine including a receiver (uplink) and transmitter (downlink) section (Auer et al., 2011; López-Pérez et al., 2022). Therefore, this power is consumed by this transmission equipment and also air conditioning units, in order to enable the RAN from cabinets at the base of the site tower (made out of aluminum sheets and polyurethane foam) (Spagnuolo et al., 2015).

Next, the emissions in the $i$th region arising from this power consumption are estimated, for carbon dioxide ($CO_2$) in Equation (14), nitrogen oxides ($NO_x$) in Equation (15), sulfur oxides ($SO_x$) in Equation (16), and particulate matter ($PM_{10}$) in Equation (17).

$$CO_{2_i} = (Energy_i \cdot Source_{it}) \cdot CO_2 Factor \tag{14}$$

$$NO_{x_i} = (Energy_i \cdot Source_{it}) \cdot NO_x Factor \tag{15}$$

$$SO_{x_i} = (Energy_i \cdot Source_{it}) \cdot SO_x Factor \qquad (16)$$

$$PM_{10_i} = (Energy_i \cdot Source_{it}) \cdot PM_{10} Factor \qquad (17)$$

Country-level carbon dioxide emission factors are utilized for the quantity of $CO_2$ released per kWh of on-grid electricity (Ritchie, Roser and Rosado, 2020) (the carbon intensity of electricity). Additionally, environmental emissions factors are also obtained from Argonne National Lab for nitrogen oxides ($NO_x$), sulfur oxides ($SO_x$), and particulate matter below 10 microns ($PM_{10}$) (Ou and Cai, 2020). Moreover, International Energy Agency (IEA) forecasts from 2020-2030 are utilized to provide information on the percentage composition of energy generation by source ($Source_{it}$), e.g., for coal, gas, renewables etc. in each annual $t$ time period up to 2030 (International Energy Agency, 2021). For clarification, we treat operational emission factors from renewable electricity production for carbon and environmental emissions to be negligible, as it common in the literature (Farquharson, Jaramillo and Samaras, 2018).

Finally, data are gathered by country from the MNO industry organization (the GSM Association) (GSMA, 2021) for the percentage of on-grid and off-grid assets. As this dataset only covers 20 countries, and 19.3% of the population in LMICs (16.3% globally), mean values are utilized based on World Bank income groups (LICs 53% on-grid, LMCs 67% on-grid, UMCs 94% on-grid). Energy generated off-grid is sourced here from diesel generators.

## 3.7. Data

The different dataset utilized within the modeling approach are presented within this section, as detailed in Table 1.

Firstly, spatial boundaries are utilized from the Database of Global Administrative Areas to provide country and sub-national (level 1 and 2) shapefiles for the analysis. Secondly, the demand model utilizes the 1 km$^2$ 2020 global population mosaic data from WorldPop (Tatem, 2017; WorldPop, 2019), along with historical subscription adoption data from GSMA Intelligence (GSMA, 2020). Thirdly, data are taken from TowerXchange continental dossiers on the number of mobile sites present in each country, and supplemented in the case of missing data with site estimates developed from OpenCelliD data. Finally, the World Energy Outlook provides forecasts of the annual energy generated by source for global regions between 2023-2030.

Table 1 Model datasets

| Model Component | Data Description | Source(s) | Reference |
|---|---|---|---|
| Boundaries | Global Identifier (GID) level 0, 1 and 2 regions | GADM | (GADM, 2019) |
| Demand | 1 km2 Population Mosaic | WorldPop | (Tatem, 2017; WorldPop, 2019) |
| Demand | Subscription adoption | GSMA | (GSMA, 2020) |
| Demand | Smartphone adoption | GSMA | (GSMA, 2020) |
| Capacity | Tower counts by country | TowerXchange | (TowerXchange, 2017, 2018b, 2018a, 2019c, 2019b, 2019a) |
| Capacity | Cell counts by country | OpenCelliD | (OpenCelliD, 2022) |
| Energy and Emissions | IEA World Energy Outlook 2020-2030 | IEA | (International Energy Agency, 2021) |

Additionally, the data values used in the wireless network simulation, are presented in Table 2, enabling the site density to area capacity lookup tables to be generated.

Table 2 Mobile system simulation parameters

| Parameter | Value | Unit |
|---|---|---|
| Antenna type | Macrocell | - |
| Frequencies (4G) | 800, 1800, 2500 | MHz |
| Frequencies (5G) | 700, 3500 | MHz |
| DL Bandwidth 4G | 10 | MHz |
| DL Bandwidth 5G | 10 (700 MHz), 40 (3500 MHz) | MHz |
| Technology | 4G, 5G | - |
| Transmission type | 2x2 MIMO (4G), 4x4 MIMO (5G) | Antennas |
| Per user capacity targets | 25 (low), 50 (baseline), 75 (high) | GB/Month |
| Transmitter power | 40 | dBm |
| Transmitter height | 30 | Meters |
| Transmitter gain | 16 | Decibels |
| Transmitter losses | 1 | Decibels |
| Receiver gain | 0 | Decibels |
| Receiver losses | 4 | Decibels |
| Receiver misc. losses | 4 | Decibels |
| Receiver height | 1.5 | Meters |
| LOS breakpoint | 500 | Meters |
| Network load | 100 | % |
| Sectorization | 3 | - |
| Shadow fading log-normal distribution | (μ, σ) = (2, 10) | dB |
| Propagation Model | Free space path loss | - |

# 4. Results

In this section the results obtained from utilizing the articulated method are presented. Firstly, the present value of the financial cost of universal mobile broadband associated with each traffic consumption and adoption scenario is reported in Figure 2**Error! Reference source not found.**, for 4G and 5G strategies using either a wireless (W) or fiber (F) backhaul.

In the baseline adoption scenario for 30 GB/month per user, a 4G strategy is estimated at $6.9 trillion when using a wireless backhaul, or $11.2 trillion when using a fiber backhaul (a 62% increase). For the same adoption scenario, 5G investment costs were lower. For example, a 5G strategy is estimated at $2.6 trillion when using a wireless backhaul, or $3.7 trillion when using a fiber backhaul (a 42% increase), thanks to 5G's higher efficiency requiring fewer sites to meet the same traffic load.

Comparing across the traffic quantities, the 4G strategy using a wireless backhaul increased to $9.7 trillion (an increase of 41%) for 75 GB/month per user, compared to a decrease to $4.3 trillion (a decrease of 38%) for 25 GB/month per user. In contrast, when comparing across the adoption scenarios, the cost for a 4G strategy using a wireless backhaul increased to $10.6 trillion (an increase of 54%) with high adoption (e.g., 4.4% CAGR), compared to a decrease to $5.2 trillion (a decrease of 25%) with the low adoption (e.g., 1.3% CAGR).

In Figure 3 the investment costs associated with mobile universal broadband are reported for different infrastructure sharing business models. The results are presented for 4G using a wireless backhaul as this will be the most common way to deliver wide-area mobile connectivity globally up to 2030.

*Figure 2 Estimates of financial cost by technology.*

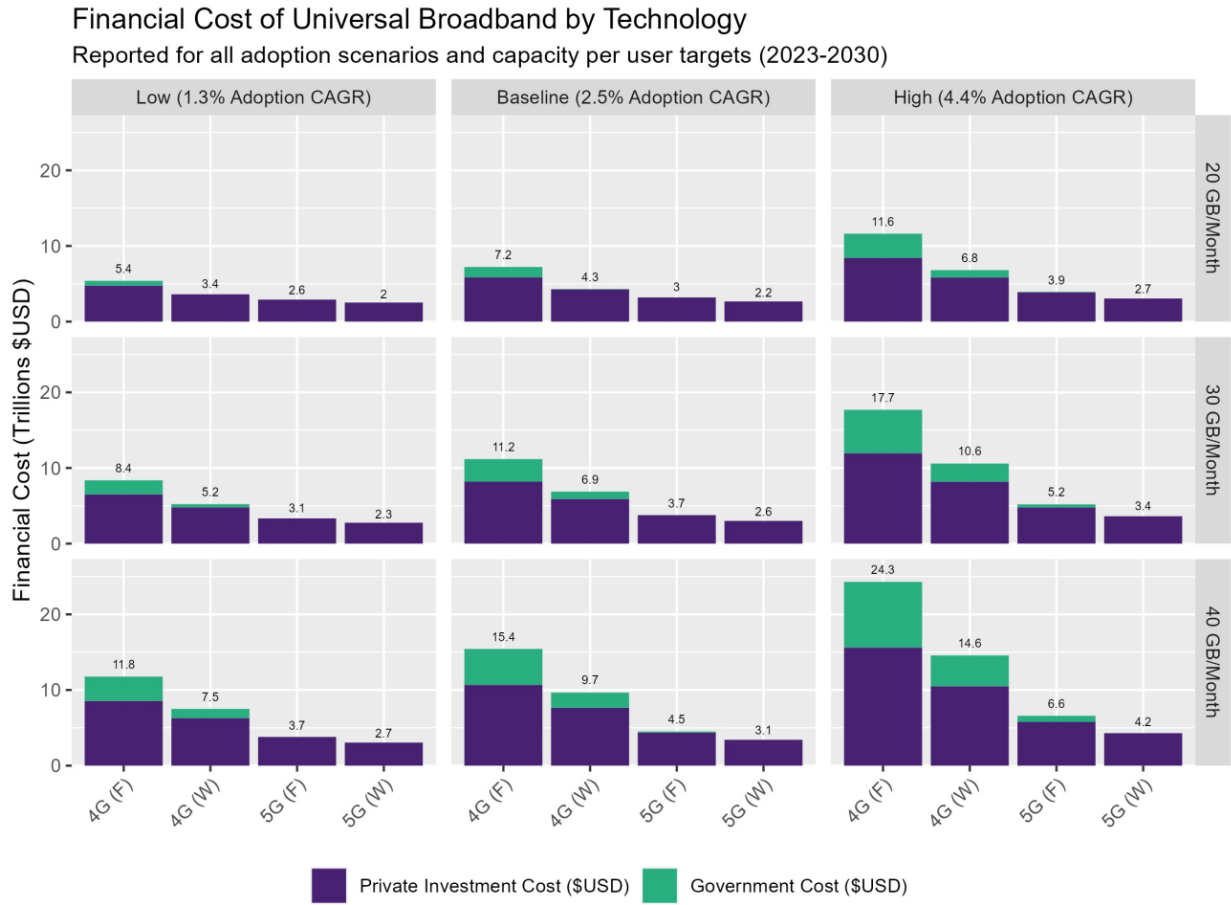

In the baseline adoption scenario with users consuming 30 GB/month, a cost saving of 39% is achieved by utilizing passive infrastructure sharing ($4.2 trillion), when compared to a business-as-usual strategy with no sharing ($6.9 trillion). The SRN ($3.5 trillion), which maintains infrastructure competition in urban and rural areas, but utilizes active sharing in rural areas, achieves a cost saving of 58% against a strategy with no sharing. In contrast, an active sharing strategy ($2.9 trillion) which implements a single shared neutral host network nationally, sees a saving of 49% when compared to a no sharing strategy. However, this may not be the most beneficial approach if preserving infrastructure competition is desired, but the result is interesting information nonetheless.

*Figure 3 Estimates of financial cost by business model.*

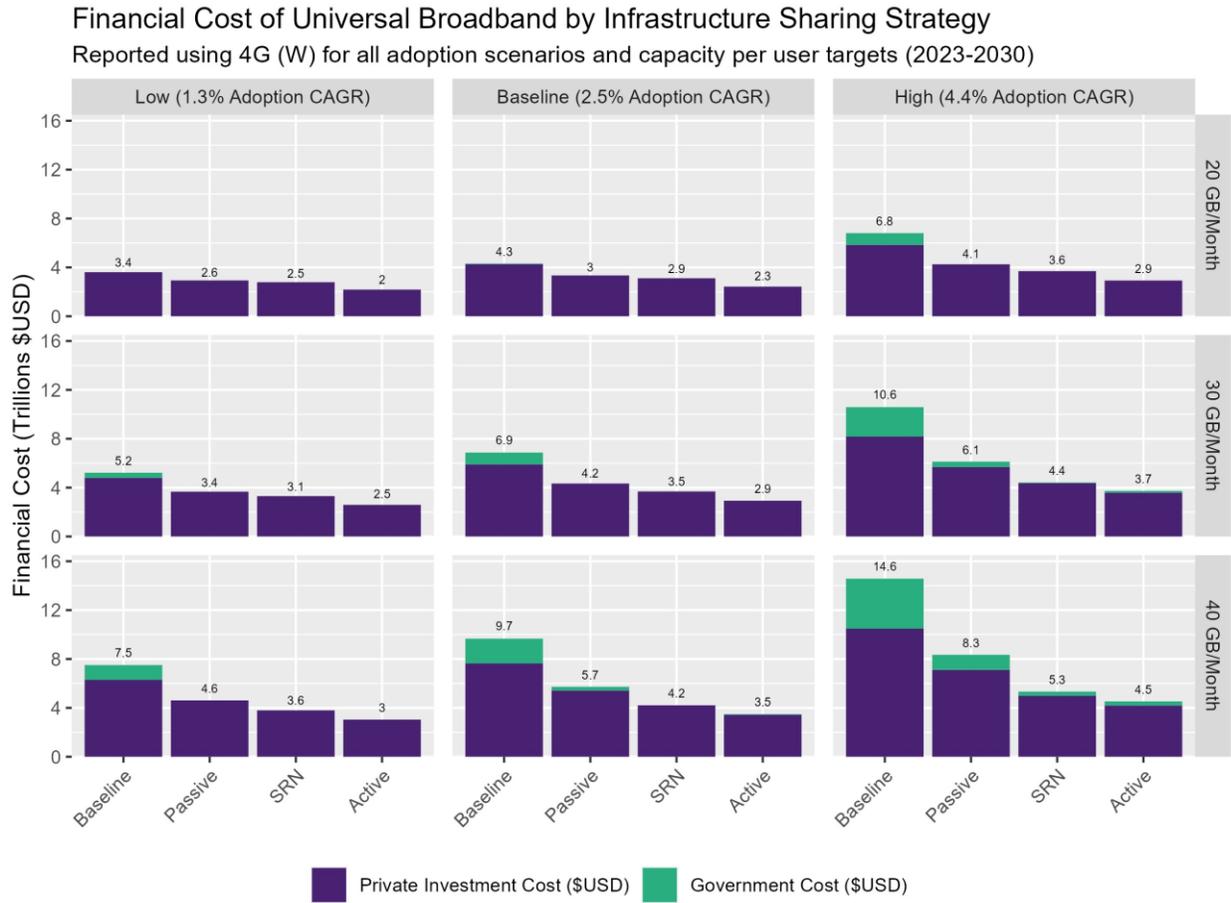

The investment costs involved in a variety of fiscal policy options are reported in Figure 4. As the modeling method adopts a user cross-subsidization approach (reallocation of excess profits to less viable areas), only marginal differences exist between changes in the fiscal and spectrum license fee options. For example, in the baseline adoption scenario where users consume 30 GB/month via 4G using a wireless backhaul, a decrease of 4% takes place when moving from the baseline ($6.9 trillion) in a low taxation strategy ($6.6 trillion). In comparison, a higher tax strategy for the same adoption and traffic quantities sees a 4% increase in cost, to a total investment of $7.2 trillion.

*Figure 4 Estimates of financial cost by policy option*

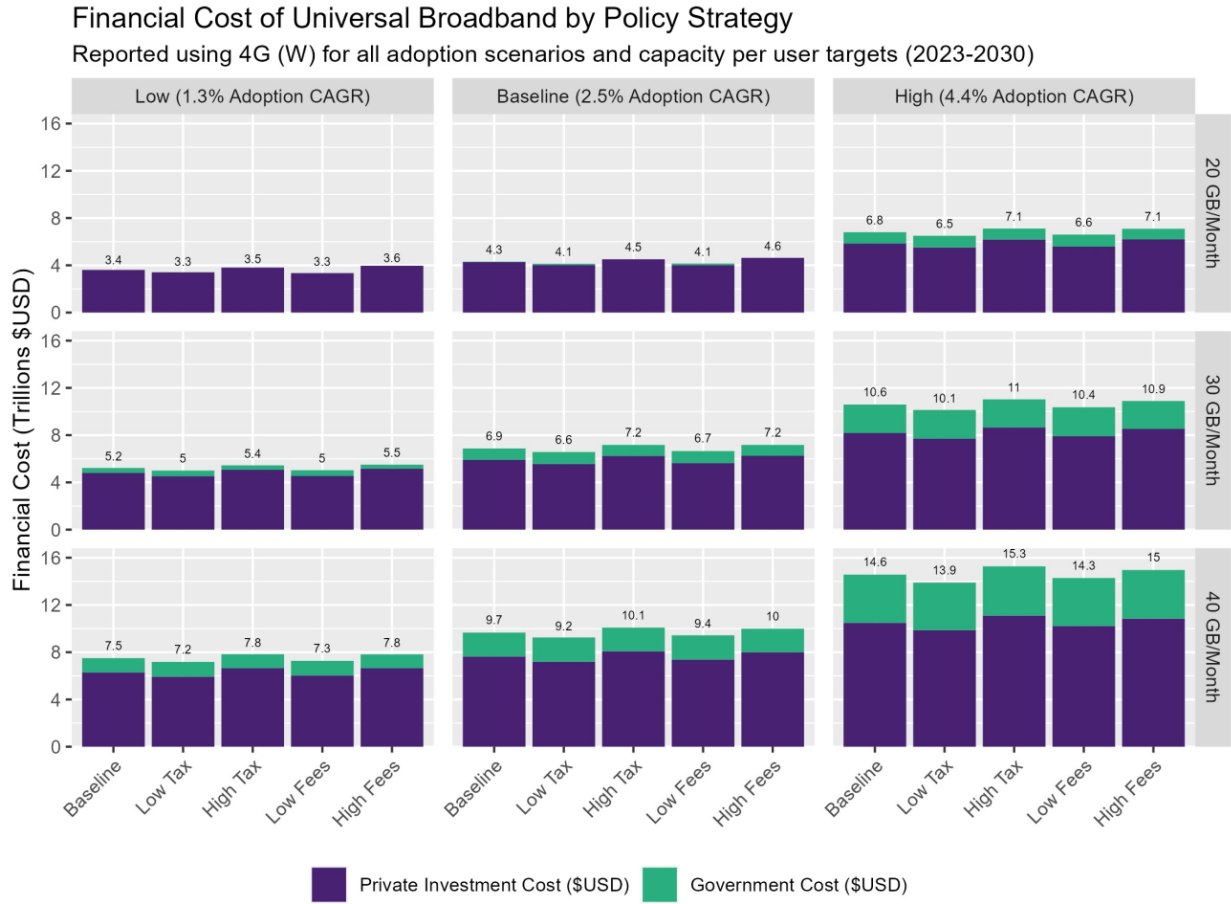

For spectrum license fee variants, marginally larger changes in total cost are estimated. For example, when using 4G with a wireless backhaul to deliver 30 GB/month under a baseline adoption scenario, the total cost decreases by 3% from $6.9 trillion in the baseline, to $6.7 trillion, when applying lower fees. However, if spectrum license fees are increased under the same adoption and traffic scenarios, the total cost increases by 4% overall, up to $7.2 trillion.

Fiscal savings are higher in those scenarios where larger investments are required. For example, in the high adoption scenario for 30 GB/month, lowering taxation leads to a 5% saving (from $10.6 trillion to $10.1 trillion). Equally, in the baseline adoption scenario for 40 GB/month, lowering spectrum fees sees a 3% saving (from $9.7 trillion to $9.4 trillion).

Delivering universal broadband is linked to increased energy consumption and $CO_2$ emissions, as additional energy-consuming assets are needed to expand network coverage. Figure 5 (A) visualizes the corresponding energy usage for different technology and capacity strategies, accounting for the forecast energy generation mix in operation up to 2030. For example, in the 30 GB per user target, approximately 83.9 Terawatt Hours (TWh) of energy is estimated to be consumed up to 2030 for a 4G strategy using a wireless backhaul. This quantity lowers to 48.6 TWh (a 42% decrease) for 20 GB/month of traffic per user and raises to 122.1 TWh (a 46% increase) for 40 GB/month of traffic per user. Generally, wireless backhaul strategies use considerably more energy than fiber backhaul strategies. Moreover, 4G is marginally more energy efficient at lower data consumption rates (e.g., 20 GB/Month), whereas 5G is more energy efficient at higher data consumption rates (e.g., 40 GB/Month).

In Figure 5 (B) the cumulative cell site $CO_2$ emissions are estimated for a 4G wireless backhaul strategy producing 5.2 Megatonnes (Mt) from the operation of assets to deliver 30 GB/month per user, up to 2030. This compared to 2.2 Mt produced by 5G with a wireless backhaul (a 58% decrease). Moreover, 5G related emissions increase less significantly than 4G when extra capacity per user is considered, thanks to the spectral efficiency improvements between these generations (as the higher capacity per site, means fewer sites are required). For example, in the 40 GB/month capacity scenario, a 4G wireless backhaul strategy sees emissions increase from 5.2 Mt up to 7.5 Mt (a 44% increase). Yet a 5G wireless backhaul strategy sees an increase in emissions from 2.2 Mt up to 3 Mt (a 36% increase).

The environmental emissions associated with each strategy are reported in Figure 6. The largest quantity of emissions released is from sulfur oxides ($SO_x$), as per Figure 6 (B). For example, for a strategy utilizing 4G with a wireless backhaul to serve 30 GB/month per user, 132.3 Kilotonnes (kt) are estimated to be released, compared to 58.8 kt for 5G with a wireless backhaul. For 4G and 5G using fiber backhaul, estimated emissions are 109.6 Mt and 55.2 Mt, respectively. These $SO_x$ emissions increase substantially at higher traffic quantities, such as for 40 GB/month per user, with a mean increase of 39%. Indeed, 4G using a wireless backhaul rises to 193.2 kt, equating to a 46% increase. At lower traffic quantities, such as 20 GB/month per user, emissions dropped on average by 36%.

*Figure 5 Estimates of energy demand and emissions by technology and traffic demand.*

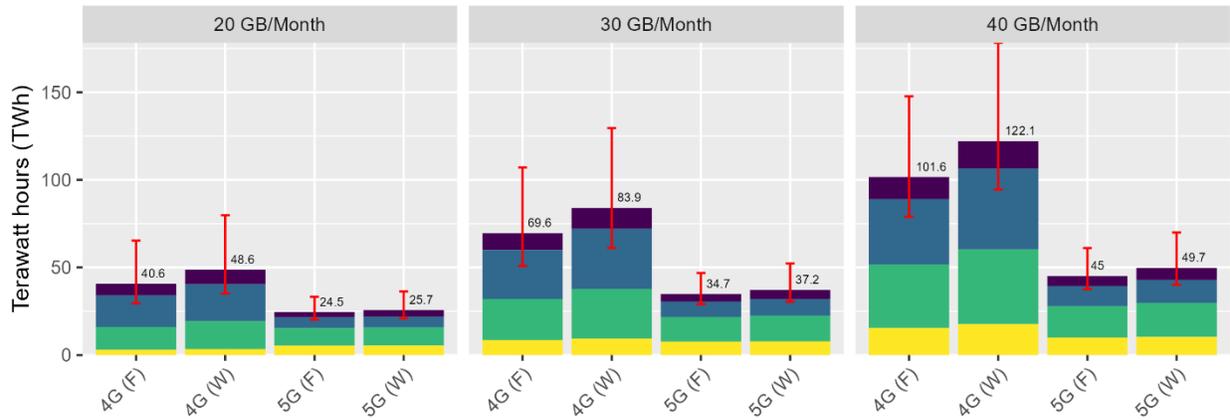

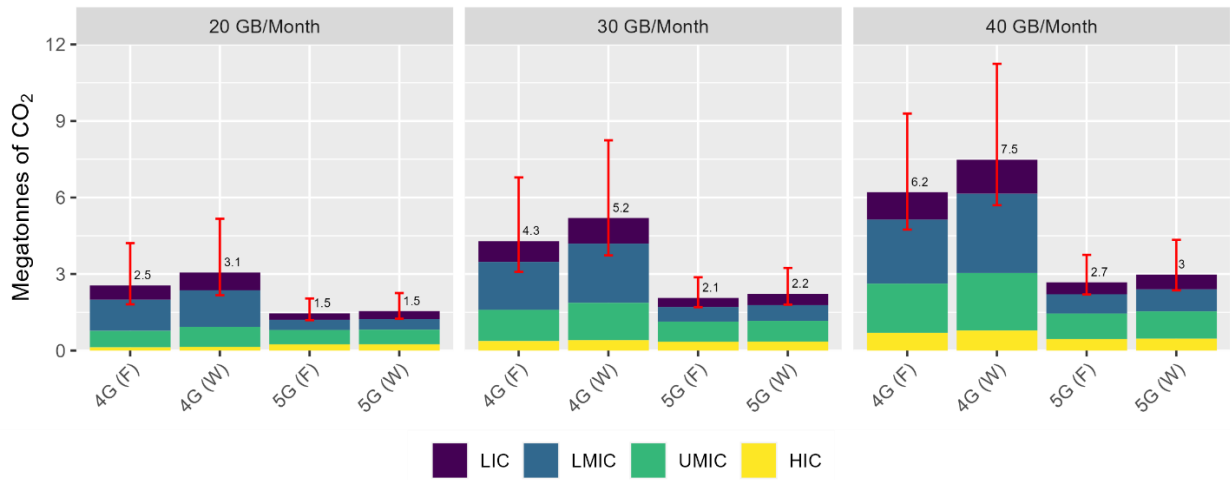

The next largest pollutant is particulate matter (PM10), where an estimated 80.8 kt is associated with a 4G strategy using a wireless backhaul, or 36 kt with a 5G strategy using a wireless backhaul, as per Figure 6 (C). Equally, a 4G strategy using a fiber backhaul is estimated to produce emissions of 66.9 kt, compared with a 5G strategy using a fiber backhaul 66.9 Mt. Under higher per user consumption levels (e.g., 40 GB/month) these values increase on average 39%, whereas they decrease on average by 37% for lower consumption levels (e.g., 20 GB/month). Only very minor quantities of nitrogen oxides (NOx) are released relative to the other environmental pollutants assessed.

*Figure 6 Estimates of environmental emissions by technology and traffic demand.*

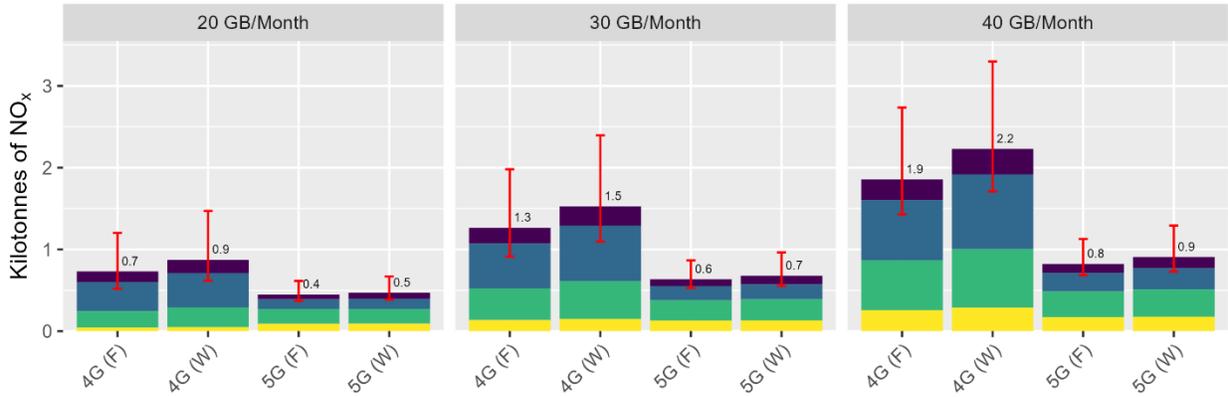

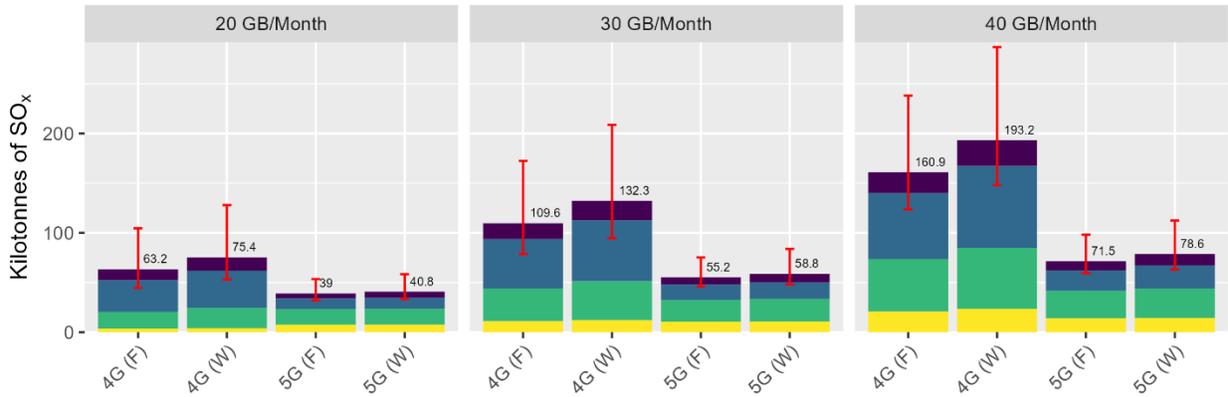

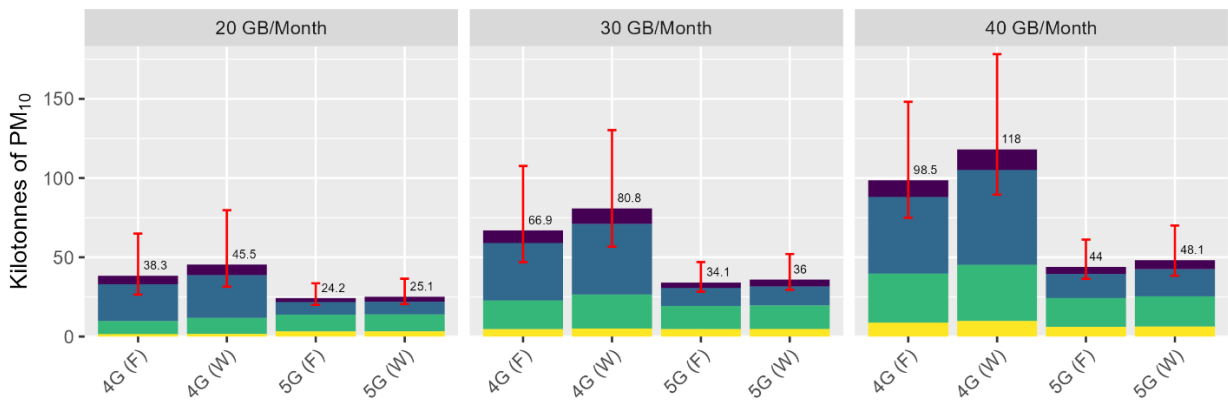

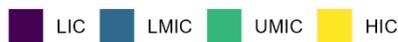

In Figure 7, the energy and emissions impact of different infrastructure sharing business models are evaluated. In Figure 7 (A), energy demand is illustrated for the baseline case where no infrastructure sharing takes place when delivering 30 GB/month. Using 4G with a wireless backhaul leads to an energy demand of 83.9 TWh, with this value remaining static for a passive sharing strategy, while seeing a 39% decrease for an active sharing strategy (to 51.5 TWh), and a 20% decrease for an SRN sharing strategy (to 66.9 TWh). On average across the technology strategies, active sharing reduced energy consumption by 26%, compared to a 9% reduction when implementing an SRN.

*Figure 7 Estimates of energy demand and emissions by business model.*

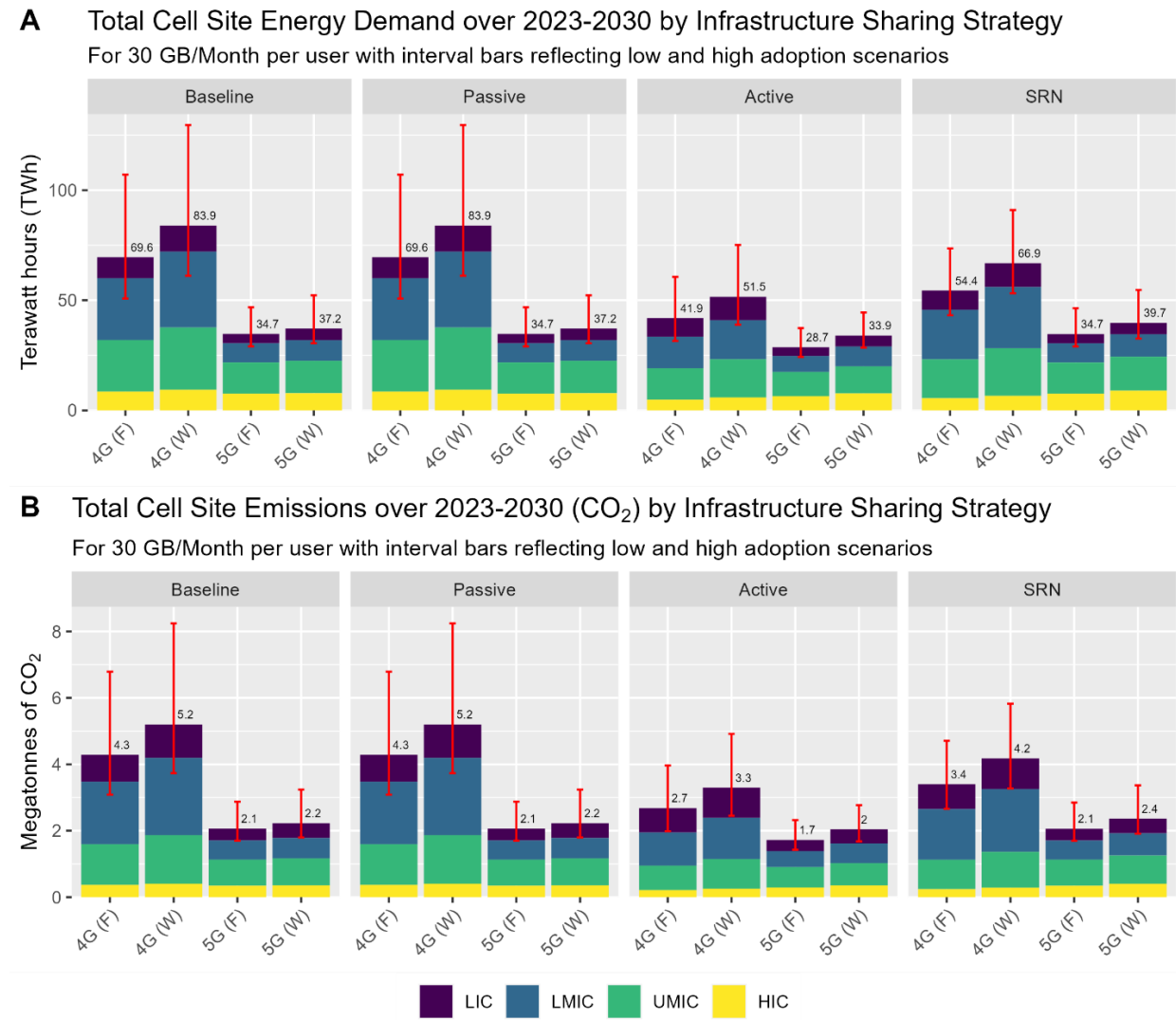

In Figure 7 (B), $CO_2$ emissions are reported for different infrastructure sharing business models when delivering 30 GB/month per user. Using 4G with a wireless backhaul produces emissions equating to 5.2 Mt, with this value remaining static for a passive sharing strategy, while seeing a 37% decrease for active and 19% decrease for SRN sharing strategies (to 3.3 and 4.3 Mt, respectively). Overall, the active sharing approach leads to a mean reduction in emissions of approximately 25% across the technology strategies, compared to a mean reduction of 8% for the SRN business model.

Similar reductions in environmental emissions are reported for the active and SRN strategies, in Figure 8. For example, in Figure 8 (B) the baseline emissions quantity for $SO_x$ at 30 GB/month is estimated to be 132.3 kt for 4G with a wireless backhaul, decreasing by 40% to 79.8 kt in an active sharing strategy, and by 21% to 104 kt in an SRN strategy. Whereas for $PM_{10}$, in Figure 8 (C) the baseline emissions for 30 GB/month via 4G with a wireless backhaul reach a total of 80.8 kt. For an active sharing strategy, this quantity decreases to 45.2 kt (a reduction of 44%), compared to an SRN strategy where this value decreases to 62 kt (a reduction of 23%).

*Figure 8 Estimates of environmental emissions by business model.*

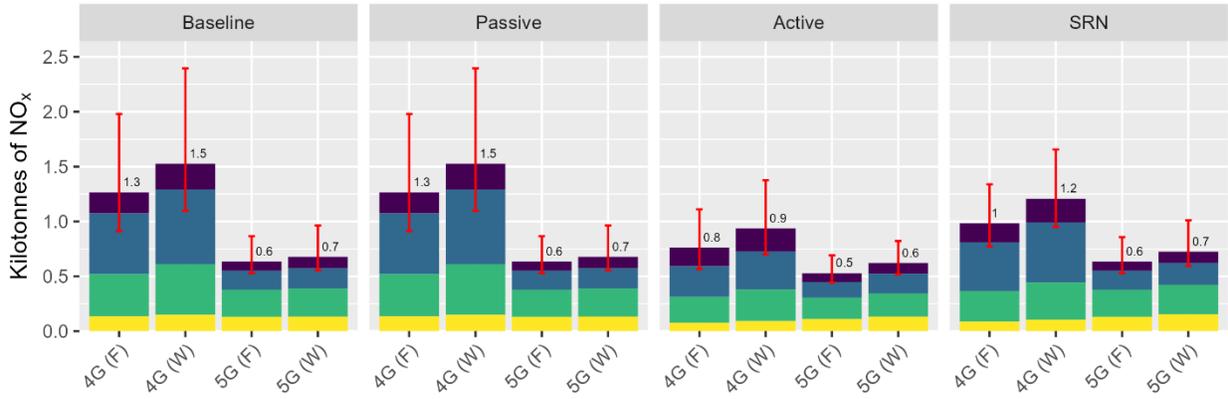

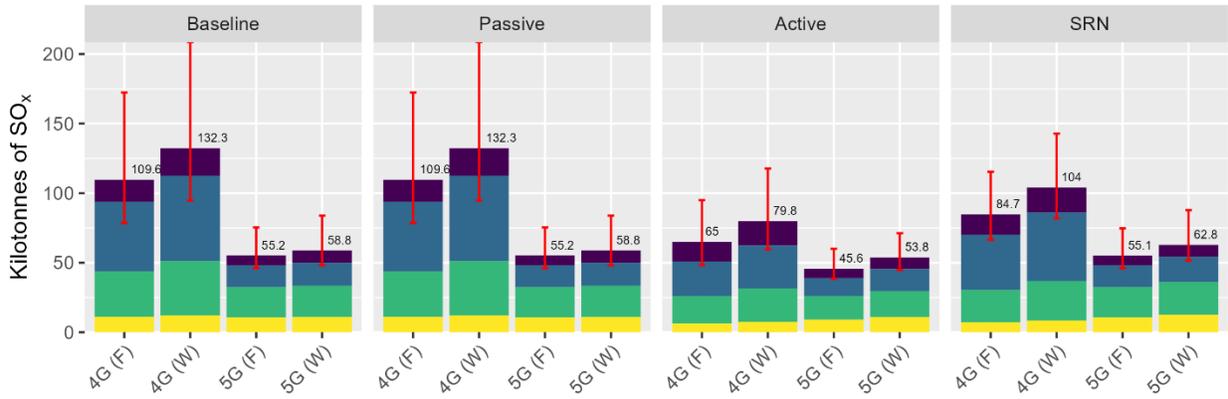

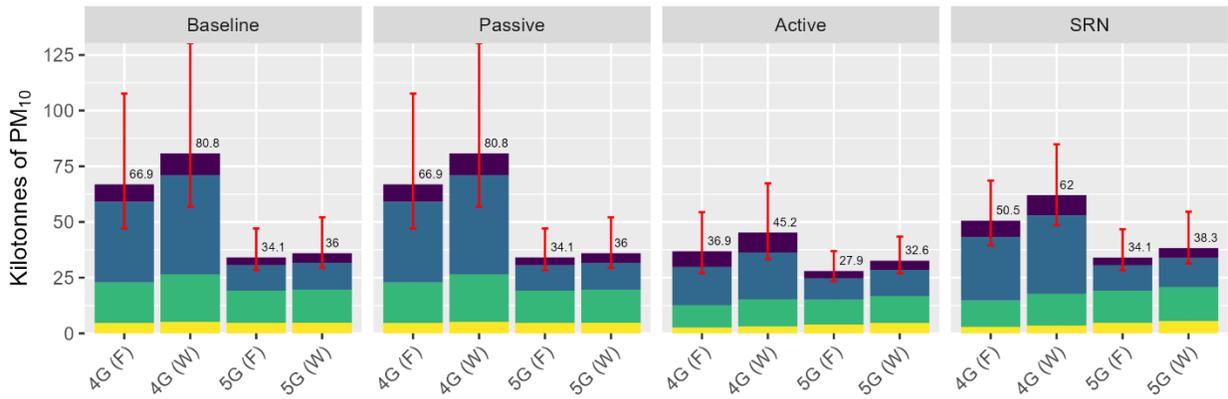

Figure 9 reports the estimated emissions benefits from converting off-grid sites powered by diesel generators to renewable power, for per user consumption quantities of 30 GB/month. The baseline for each technology is reported, representing the business-as-usual case, whereas the renewables comparison estimates the total emissions from using renewable power options.

*Figure 9 Estimates of saved environmental emissions for renewable off-grid sites*

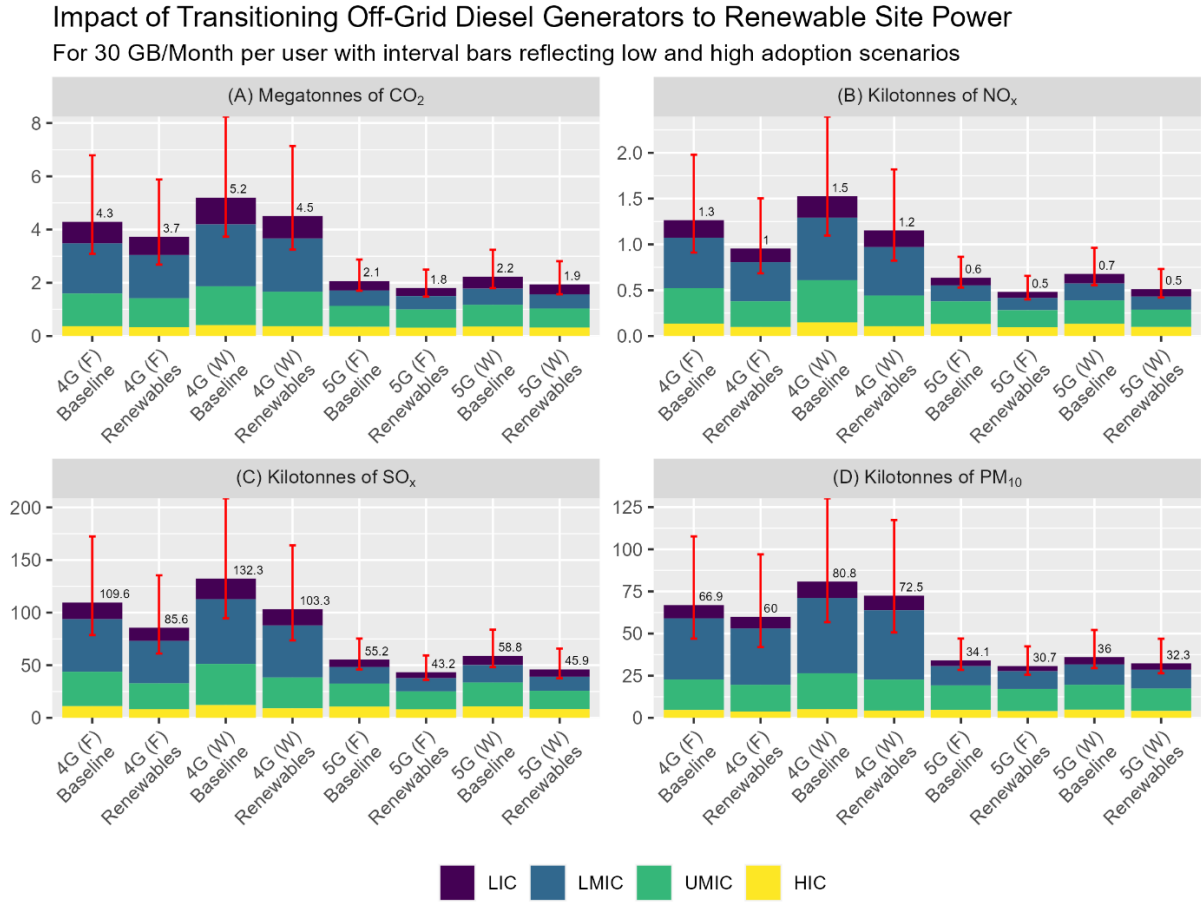

In Figure 9 (A) the mean saving in $CO_2$ emissions across all strategies equates to 13%. For example, 4G using a wireless backhaul is estimated at 5.2 Mt, compared to a renewable energy emissions quantity of 4.5 Mt. Moreover, in Figure 9 (C) emission reductions for $SO_x$ are reported by strategy, with a mean saving of 22%, whereby using 4G with a wireless backhaul the estimated quantity of $SO_x$ decreases from 132.3 kt down to 103.3 kt. Finally, for Figure 9 (D) the $PM_{10}$ emissions from universal mobile broadband could be reduced by on average 10% across the different technology strategies explored here, for example, with 4G using a wireless backhaul decreasing from 80.8 kt down to 72.5 kt.

## 5. Discussion

This discussion will review the results presented in the previous section, in relation to the four main research questions identified in the paper introduction.

*What are the financial costs of universal mobile broadband by technology, business model and policy option?*

A wide range of technology strategies, adoption scenarios and monthly traffic quantities were explored, to estimate the financial cost of universal mobile broadband in present value terms.

When examining the costs of universal mobile broadband, a 4G strategy targeting 30 GB/month in the baseline adoption scenario is estimated at $6.9 trillion when using a wireless backhaul, or $11.2 trillion when using a fiber backhaul (an increase of 62%). In contrast, 5G investment costs were lower in the same adoption scenario to deliver 30 GB/month, thanks to having improved spectral efficiency. Approximately $2.6 trillion is estimated for a 5G strategy when using a wireless backhaul or $3.7 trillion when using a fiber backhaul (an increase of 42%), to deliver 30 GB/month. On average, using a more superior fiber backhaul approach is 52% more expensive when compared to strategies which utilize a wireless backhaul approach.

The results indicate that the changes explored in the monthly traffic have a substantial impact on cost when compared to varying the adoption scenario. For example, on average across all technology strategies, a cost increase of 30% takes place when switching from 30 GB/month, up to 40 GB/month, along with a cost decrease of 27% when switching to 20 GB/month (e.g., from $6.9 Tn, up to $9.7 Tn, or down to $4.3 Tn for 4G with a wireless backhaul). In comparison, a 46% increase took place when the adoption was changed from the baseline to high (with a 4.4% CAGR), along with a decrease of 19% under the low adoption scenario (e.g., 1.3% CAGR) (e.g., from $6.9 Tn, up to $10.6 Tn, or down to $5.2 Tn for 4G with a wireless backhaul).

The challenge here is that 4G is a lot less cost efficient for the rather ambitious data capacities tested. But with 5G the key issue is that many unconnected or underserved users may not be able to afford the higher handset device cost. Therefore, there are key trade-offs which will have to be made to provide universal mobile broadband, as even 20 GB/month may be too high to provide in some areas using 4G. Certainly for providing basic mobile broadband for smaller traffic quantities, 4G is most likely the best immediate choice, as supported in the literature [cite Ericcson Mobility report on devices and 4G being dominant until 2028]. But for those countries which are targeting much higher data usage e.g., >=20 GB/Month, it may be that 5G is the better choice for providing to universal mobile broadband, as a much more spectrally efficient technology (although handset subsidies may well be something to consider). We also need to see what impact Reduced Capability (RedCap) 5G NR devices have on reducing handset costs, following standardization in 3GPP Release 17.

Most noticeable in the analysis is the cost savings achieve by utilizing different infrastructure sharing business models. For example, when examining a 4G wireless backhaul strategy for 30 GB/month under baseline adoption, the passive infrastructure sharing strategy delivers a cost saving of 39% (to $4.2 trillion), when compared to a business-as-usual strategy with no sharing (totaling $6.9 trillion). Moreover, an active sharing approach achieves a large saving of 58% (down to $2.9 trillion) when implementing a single shared neutral host network nationally when compared to a no sharing strategy. The caveat to this saving is that many decision makers in telecommunication regulators would be unlikely to relinquish the positive benefits of infrastructure competition. Therefore, the savings delivered by an SRN are more attractive even if they are marginally less. For example, a 49% saving is estimated ($3.5 trillion), when compared to the baseline where no sharing takes place. In this approach, market-based infrastructure competition is preserved in urban and suburban locations, with a single neutral host network deployed only in less viable rural areas.

Considerably smaller savings were estimated when examining fiscal policy options, which partially arises from the method being based on a user cross-subsidization approach (akin to specifying a national coverage obligation to force the reallocation of excess profits to unviable locations). Under the baseline adoption scenario for a 4G wireless backhaul approach, changes in taxation produced cost variation around the baseline of between 3-4% when users consume 30 GB/month. In contrast, spectrum license fees changes also led to variation around the baseline financial cost of approximately 3-4%.

*What are the energy and environmental impacts of universal mobile broadband?*

Although mobile broadband is generally considered as an enabler of carbon emissions strategies, the operation of new and existing cellular sites require energy and therefore have an environmental impact. Universal broadband strategies lead to an increase in energy consumption and emissions because new sites need to be deployed to provide coverage in previously unconnected or poorly served regions. The modeling produced in this paper estimates that for a 30 GB/month target with baseline adoption, approximately 83.9 TWh of energy is cumulatively consumed up to 2030, when using a 4G strategy with a wireless backhaul. This quantity decreases by 42% to 48.6 TWh for 20 GB/month, whereas it increased by 46% to 122.1 TWh for 40 GB/month. Fiber backhaul strategies consume considerably less energy than wireless backhaul strategies.

In terms of cumulative $CO_2$ emissions to 2030, a 4G wireless backhaul strategy produces approximately 5.2 Mt from the operation of all assets, capable of providing 30 GB/month. When using 5G with a wireless backhaul a decrease of 58% is achieved (to 2.2 Mt). A 4G wireless backhaul strategy sees emissions increase by 44% (up to 7.5 Mt) when moving from 30 GB/month to 40 GB/month. Whereas when comparing the same traffic increase for 5G with a wireless backhaul, only a 36% increase in emissions is estimated (from 2.2 Mt up to 3 Mt).

A variety of environmental emissions were quantified for each strategy explored. Across the different adoption scenarios and traffic quantities, the largest cumulative emissions up to 2030 take place from $SO_x$, followed by $PM_{10}$ and $NO_x$. For example, in the 4G wireless backhaul technology strategy, 132.3 kt of $SO_x$, 80.8 kt of $PM_{10}$ and 1.5 kt of $NO_x$ emissions are estimated when serving 30 GB/month under baseline adoption.

*How do different business model options affect the energy and environmental impacts of universal mobile broadband?*

Reducing the amount of active infrastructure which needs to be built, leads to lower energy consumption and therefore a reduction in emissions. For example, in the baseline adoption scenario for 30 GB/month, a 4G wireless backhaul strategy consumes a cumulative 83.9 TWh of energy. However, implementing an active sharing strategy sees a decline to 51.5 TWh (a decrease of 39%), compared to an SRN strategy which falls to 66.9 TWh (a decrease of 20%). Across the scenarios and technology strategies, the SRN business model saw a mean decrease of 9% in energy consumption, compared to a mean decrease of 26% for the active infrastructure sharing business model.

With energy consumption dropping substantially under business models which utilize active sharing, so too do associated emissions. For example, a nationally homogenous active sharing approach led to a mean decrease of 37% in $CO_2$ emissions across the technology strategies, with a mean decrease for the SRN business model of 19%. So rather than 5.2 Mt of cumulative emissions for 4G using a wireless backhaul, an SRN approach would incur as low as 4.2 Mt of $CO_2$ (under baseline adoption for 30 GB/month). Similar decreases in environmental emissions are estimated for a wireless backhaul 4G approach, with SOx declining by 40% under active sharing (to 79.8 kt), and 21% for an SRN strategy (to 104 kt). These findings can be compared to $PM_{10}$ emissions, which also experienced a decrease by 44% (45.2 kt) for active sharing, and 23% (62 kt) for the SRN strategy.

*To what extent do environmental benefits arise from shifting off-grid diesel sites to renewable power?*

Subject to the modeling exercise carried out in this paper, emissions savings are found by converting off-grid diesel generators to utilizing renewable power options. For example, for 30 GB/month under baseline adoption, $CO_2$ emissions could be reduced on average by 13% across all 4G and 5G strategies explored. Similar reductions were found for $SO_x$ emissions with an average reduction of 22%, and for $PM_{10}$ emissions with an average of 10% decrease across the explored technology strategies.

Two key uncertainties existing in the results presented for this research question. Firstly, assumptions are used for the number of off-grid sites utilizing diesel generators. For example, on the former point, there is a lack of country information on the number of diesel generators in use, meaning this analysis had to draw on a sample of countries provided by the GSMA climate tech dataset. How generator usage varies across all low and middle-income countries is know fully known or understood. Indeed, many network operators may be unlikely to disclose their full usage of this low-cost power option for commercial reasons. Secondly, there may be practical issues which affect the deployment of reliable renewable power options for some locations. For example, photovoltaic technologies function well ±30° latitude, but at higher latitudes the number of hours of daily sunlight can be highly variable, especially in the winter months. Wind is an alternative power option which may work well for some locations. However, this approach also can be affected by intermittency. Future research needs to example the practical implications of the deployment of these technologies.

# 6. Conclusions

The aim of this paper has been to assess the sustainability implications of providing universal mobile broadband across the globe. This was achieved by firstly considering the financial costs involved in providing technology, business model and policy options. Secondly, the energy consumption and environmental emissions impacts associated with universal mobile broadband strategies. Thirdly, the impact of utilizing different business model options on the associated energy consumption and environmental emissions impacts from universal mobile broadband. And finally, the potential environmental benefits of shifting off-grid diesel-power sites onto renewable energy generating technologies.

This paper identifies a set of key findings. The deployment of 5G is generally cheaper to deploy than 4G, thanks to improved spectral efficiency, but this would pose problems for universal broadband as devices with newer chipsets are generally more expensive. Moreover, a wireless backhaul is generally cheaper than using a fixed fiber connection, although this approach uses more energy and therefore has a larger operational environmental impact.

One way to reduce energy demand and emissions is via business model innovation (focusing on sharing infrastructure assets) which helps to reduce infrastructure duplication. However, decision makers need to consider the competition implications of infrastructure sharing. Finally, utilizing renewable energy strategies for off-grid sites, can help to reduce energy demand (by ~13%) and thus emissions (by 10-22% depending on the emissions type), by reducing dependence on diesel generators.

The key contribution to the literature is in providing these sustainability analytics for future universal mobile strategies, helping to emphasize trade-offs in technology, business model and policy option choices. Future research should consider the energy and emissions implications of producing electronic equipment, concrete, steel and other materials required to provide broadband services.

# Bibliography

Abrardi, L. and Sabatino, L. (2023) 'Ultra-broadband investment and economic resilience: Evidence from the Covid-19 pandemic', *Telecommunications Policy*, 47(2), p. 102480. Available at: https://doi.org/10.1016/j.telpol.2022.102480.

Ahmad, T. *et al.* (2015) 'Solar vs diesel: Where to draw the line for cell towers?', in *Proceedings of the Seventh International Conference on Information and Communication Technologies and Development*, pp. 1–11.

Amole, A.O. *et al.* (2021) 'Comparative Analysis of Techno-Environmental Design of Wind and Solar Energy for Sustainable Telecommunications Systems in Different Regions of Nigeria', *International Journal of Renewable Energy Research (IJRER)*, 11(4), pp. 1776–1792.

Andrae, A.S.G. and Edler, T. (2015) 'On Global Electricity Usage of Communication Technology: Trends to 2030', *Challenges*, 6(1), pp. 117–157. Available at: https://doi.org/10.3390/challe6010117.

Auer, G. *et al.* (2011) 'How much energy is needed to run a wireless network?', *IEEE Wireless Communications*, 18(5), pp. 40–49. Available at: https://doi.org/10.1109/MWC.2011.6056691.

Baidas, M.W. *et al.* (2022) 'Renewable-Energy-Powered Cellular Base-Stations in Kuwait's Rural Areas', *Energies*, 15(7), p. 2334. Available at: https://doi.org/10.3390/en15072334.

Belkhir, L. and Elmeligi, A. (2018) 'Assessing ICT global emissions footprint: Trends to 2040 & recommendations', *Journal of Cleaner Production*, 177, pp. 448–463. Available at: https://doi.org/10.1016/j.jclepro.2017.12.239.

Chen, Y.P. *et al.* (2023) 'Crowdsourced data indicates broadband has a positive impact on local business creation', *Telematics and Informatics*, 84, p. 102035. Available at: https://doi.org/10.1016/j.tele.2023.102035.

Clément, L.-P.P.-V.P., Jacquemotte, Q.E.S. and Hilty, L.M. (2020) 'Sources of variation in life cycle assessments of smartphones and tablet computers', *Environmental Impact Assessment Review*, 84, p. 106416. Available at: https://doi.org/10.1016/j.eiar.2020.106416.

Crawford, M.M. (2019) 'A comprehensive scenario intervention typology', *Technological Forecasting and Social Change*, 149, p. 119748. Available at: https://doi.org/10.1016/j.techfore.2019.119748.

Deller, S., Whitacre, B. and Conroy, T. (2022) 'Rural broadband speeds and business startup rates', *American Journal of Agricultural Economics*, 104(3), pp. 999–1025. Available at: https://doi.org/10.1111/ajae.12259.

Elzen, B. *et al.* (2002) 'Socio-Technical Scenarios as a tool for Transition Policy An example from the traffic and transport domain', *Network*, (June), pp. 23–26.

Ericsson (2023) *Ericsson Mobility Report*. Stockholm, Sweden: Ericsson. Available at: https://www.ericsson.com/en/reports-and-papers/mobility-report (Accessed: 24 July 2023).

ETSI (2018) *5G; NR; Physical Layer Procedures for Data (3GPP TS 38.214 Version 15.3.0 Release 15)*. Valbonne, France: European Telecommunications Standards Institute. Available at: https://www.etsi.org/deliver/etsi_ts/138200_138299/138214/15.03.00_60/ts_138214v150300p.pdf.

Farquharson, D., Jaramillo, P. and Samaras, C. (2018) 'Sustainability implications of electricity outages in sub-Saharan Africa', *Nature Sustainability*, 1(10), pp. 589–597. Available at: https://doi.org/10.1038/s41893-018-0151-8.

Freitag, C. *et al.* (2021) 'The real climate and transformative impact of ICT: A critique of estimates, trends, and regulations', *Patterns*, 2(9), p. 100340. Available at: https://doi.org/10.1016/j.patter.2021.100340.

Frias, Z., González-Valderrama, C. and Pérez Martínez, J. (2017) 'Assessment of spectrum value: The case of a second digital dividend in Europe', *Telecommunications Policy*, 41(5), pp. 518–532. Available at: https://doi.org/10.1016/j.telpol.2016.12.008.

Frias, Z., Mendo, L. and Oughton, E.J. (2020) 'How Does Spectrum Affect Mobile Network Deployments? Empirical Analysis Using Crowdsourced Big Data', *IEEE Access*, 8, pp. 190812–190821. Available at: https://doi.org/10.1109/ACCESS.2020.3031963.

Frith, D. and Tapinos, E. (2020) 'Opening the "black box" of scenario planning through realist synthesis', *Technological Forecasting and Social Change*, 151, p. 119801. Available at: https://doi.org/10.1016/j.techfore.2019.119801.

GADM (2019) *Global Administrative Areas Database (Version 3.6)*. Available at: https://gadm.org/ (Accessed: 11 July 2019).

Gordon, A.V. (2020) 'Limits and longevity: A model for scenarios that influence the future', *Technological Forecasting and Social Change*, 151, p. 119851. Available at: https://doi.org/10.1016/j.techfore.2019.119851.

GSMA (2020) *GSMA Intelligence Global Data*. Available at: https://www.gsmaintelligence.com/ (Accessed: 5 February 2020).

GSMA (2021) 'ClimateTech', *Mobile for Development*. Available at: https://www.gsma.com/mobilefordevelopment/renewable-energy-dashboard/ (Accessed: 27 July 2021).

Gupta, U. *et al.* (2022) 'Chasing Carbon: The Elusive Environmental Footprint of Computing', *IEEE Micro*, 42(4), pp. 37–47. Available at: https://doi.org/10.1109/MM.2022.3163226.

Haldar, A. and Sethi, N. (2022) 'Environmental effects of Information and Communication Technology - Exploring the roles of renewable energy, innovation, trade and financial development', *Renewable and Sustainable Energy Reviews*, 153, p. 111754. Available at: https://doi.org/10.1016/j.rser.2021.111754.

Hall, J.W. *et al.* (2016) *The future of national infrastructure: A system-of-systems approach*. Cambridge University Press.

Hall, J.W. *et al.* (2017) 'Strategic analysis of the future of national infrastructure', *Proceedings of the Institution of Civil Engineers - Civil Engineering*, 170(1), pp. 39–47. Available at: https://doi.org/10.1680/jcien.16.00018.

Hasbi, M. (2020) 'Impact of very high-speed broadband on company creation and entrepreneurship: Empirical Evidence', *Telecommunications Policy*, 44(3), p. 101873. Available at: https://doi.org/10.1016/j.telpol.2019.101873.

Hossain, M.S. *et al.* (2020) 'Renewable Energy-Aware Sustainable Cellular Networks with Load Balancing and Energy-Sharing Technique', *Sustainability*, 12(22), p. 9340. Available at: https://doi.org/10.3390/su12229340.

Hossain, M.S. *et al.* (2021) 'Techno-Economic Analysis of the Hybrid Solar PV/H/Fuel Cell Based Supply Scheme for Green Mobile Communication', *Sustainability*, 13(22), p. 12508. Available at: https://doi.org/10.3390/su132212508.

International Energy Agency (2021) *World Energy Outlook 2021*, *IEA*. Available at: https://www.iea.org/reports/world-energy-outlook-2021 (Accessed: 10 September 2023).

International Monetary Fund (2021) *World Economic Outlook (October 2020) - Inflation rate, average consumer prices*. Available at: https://www.imf.org/external/datamapper/PCPIPCH@WEO (Accessed: 3 March 2021).

International Telecommunication Union (2020) *Connecting Humanity*. Geneva, Switzerland: International Telecommunication Union. Available at: https://www.itu.int/en/myitu/Publications/2020/08/31/08/38/Connecting-Humanity (Accessed: 9 March 2021).

International Telecommunication Union and United Nations Educational, Scientific and Cultural Organization (2022) *The State of Broadband 2022*. Geneva, Switzerland: ITU/UNESCO. Available at: https://www.broadbandcommission.org/publication/state-of-broadband-2022/ (Accessed: 20 February 2023).

Islam, K.Z. *et al.* (2023) 'Renewable Energy-Based Energy-Efficient Off-Grid Base Stations for Heterogeneous Network', *Energies*, 16(1), p. 169. Available at: https://doi.org/10.3390/en16010169.

Isley, C. and Low, S.A. (2022) 'Broadband adoption and availability: Impacts on rural employment during COVID-19', *Telecommunications Policy*, 46(7), p. 102310. Available at: https://doi.org/10.1016/j.telpol.2022.102310.

ITU-R (2017) *Monte Carlo simulation methodology for the use in sharing and compatibility studies between different radio services or systems (Report ITU-R SM.2028-2)*. Report ITU-R SM.2028-2. Geneva, Switzerland: International Telecommunication Union.

Jahid, A. *et al.* (2021) 'Energy Efficient Throughput Aware Traffic Load Balancing in Green Cellular Networks', *IEEE Access*, 9, pp. 90587–90602. Available at: https://doi.org/10.1109/ACCESS.2021.3091499.

Jefferson, M. (2020) 'Scenario planning: Evidence to counter "Black box" claims.', *Technological Forecasting and Social Change*, 158, p. 120156. Available at: https://doi.org/10.1016/j.techfore.2020.120156.

Jones, N. (2018) 'How to stop data centres from gobbling up the world's electricity', *Nature*, 561(7722), pp. 163–167.

Kaack, L.H. *et al.* (2022) 'Aligning artificial intelligence with climate change mitigation', *Nature Climate Change*, 12(6), pp. 518–527. Available at: https://doi.org/10.1038/s41558-022-01377-7.

Koratagere Anantha Kumar, S. and Oughton, E.J. (2023) 'Techno-economic assessment of 5G infrastructure sharing business models in rural areas', *Frontiers in Computer Science*, 5. Available at: https://www.frontiersin.org/articles/10.3389/fcomp.2023.1191853 (Accessed: 31 October 2023).

Kumar, S.K.A. and Oughton, E.J. (2023) 'Infrastructure Sharing Strategies for Wireless Broadband', *IEEE Communications Magazine*, 61(7), pp. 46–52. Available at: https://doi.org/10.1109/MCOM.005.2200698.

Kusuma, J. *et al.* (2022) *Smart off-grid power for rural connectivity: Results of a field study examining the application of smart off-grid power to provide cost-effective, reliable telecommunications services in rural regions of Peru*. Wakefield, MA: Telecom Infra Project. Available at: https://telecominfraproject.com/wp-content/uploads/Smart-Off-Grid-Power-for-Rural-Connectivity-White-Paper.pdf?fbclid=IwAR26GNnj_zDVgRiHrRbK2QoyoppbVMmjkGRKvVjhxJYfFDTRECUHt-lCAwM.

Lange, S., Pohl, J. and Santarius, T. (2020) 'Digitalization and energy consumption. Does ICT reduce energy demand?', *Ecological Economics*, 176, p. 106760. Available at: https://doi.org/10.1016/j.ecolecon.2020.106760.

López-Pérez, D. *et al.* (2022) 'A Survey on 5G Radio Access Network Energy Efficiency: Massive MIMO, Lean Carrier Design, Sleep Modes, and Machine Learning', *IEEE Communications Surveys & Tutorials*, 24(1), pp. 653–697. Available at: https://doi.org/10.1109/COMST.2022.3142532.

Lundén, D. *et al.* (2022) 'Electricity Consumption and Operational Carbon Emissions of European Telecom Network Operators', *Sustainability*, 14(5), p. 2637. Available at: https://doi.org/10.3390/su14052637.

Mac Domhnaill, C., Mohan, G. and McCoy, S. (2021) 'Home broadband and student engagement during COVID-19 emergency remote teaching', *Distance Education*, 42(4), pp. 465–493. Available at: https://doi.org/10.1080/01587919.2021.1986372.

Malmodin, J. and Lundén, D. (2018) 'The Energy and Carbon Footprint of the Global ICT and E&M Sectors 2010–2015', *Sustainability*, 10(9), p. 3027. Available at: https://doi.org/10.3390/su10093027.

Masanet, E. *et al.* (2020) 'Recalibrating global data center energy-use estimates', *Science*, 367(6481), pp. 984–986. Available at: https://doi.org/10.1126/science.aba3758.

Mogensen, P. *et al.* (2007) 'LTE Capacity Compared to the Shannon Bound', in *2007 IEEE 65th Vehicular Technology Conference - VTC2007-Spring*, pp. 1234–1238. Available at: https://doi.org/10.1109/VETECS.2007.260.

Mytton, D. and Ashtine, M. (2022) 'Sources of data center energy estimates: A comprehensive review', *Joule*, 6(9), pp. 2032–2056. Available at: https://doi.org/10.1016/j.joule.2022.07.011.

Ndubuisi, G., Otioma, C. and Tetteh, G.K. (2021) 'Digital infrastructure and employment in services: Evidence from Sub-Saharan African countries', *Telecommunications Policy*, 45(8), p. 102153. Available at: https://doi.org/10.1016/j.telpol.2021.102153.

Ofcom (2018) *Mobile call termination market review 2018-21: Final statement – Annexes 1 - 15*. London: Ofcom. Available at: https://www.ofcom.org.uk/__data/assets/pdf_file/0022/112459/MCT-review-statement-annexes-115.pdf (Accessed: 5 June 2019).

OpenCelliD (2022) *OpenCelliD - Largest Open Database of Cell Towers & Geolocation - by Unwired Labs*, *OpenCelliD - Largest Open Database of Cell Towers & Geolocation - by Unwired Labs*. Available at: https://opencellid.org/#zoom=16&lat=37.77889&lon=-122.41942 (Accessed: 9 August 2022).

Ou, L. and Cai, H. (2020) *Update of Emission Factors of Greenhouse Gases and Criteria Air Pollutants, and Generation Efficiencies of the US Electricity Generation Sector*. Argonne National Lab.(ANL), Argonne, IL (United States).

Oughton, E.J. *et al.* (2019) 'Stochastic Counterfactual Risk Analysis for the Vulnerability Assessment of Cyber-Physical Attacks on Electricity Distribution Infrastructure Networks', *Risk Analysis*, 39(9), pp. 2012–2031. Available at: https://doi.org/10.1111/risa.13291.

Oughton, E.J. *et al.* (2022) 'Policy choices can help keep 4G and 5G universal broadband affordable', *Technological Forecasting and Social Change*, 176, p. 121409. Available at: https://doi.org/10.1016/j.techfore.2021.121409.

Oughton, E.J. (2023) 'Policy options for broadband infrastructure strategies: A simulation model for affordable universal broadband in Africa', *Telematics and Informatics*, 76, p. 101908. Available at: https://doi.org/10.1016/j.tele.2022.101908.

Oughton, E.J., Amaglobeli, D. and Moszoro, M. (2023) 'What would it cost to connect the unconnected? Estimating global universal broadband infrastructure investment', *Telecommunications Policy*, p. 102670. Available at: https://doi.org/10.1016/j.telpol.2023.102670.

Oughton, E.J., Boch, E. and Kusuma, J. (2022) 'Engineering-Economic Evaluation of Diffractive NLOS Backhaul (e3nb): A Techno-economic Model for 3D Wireless Backhaul Assessment', *IEEE Access*, 10, pp. 3430–3446. Available at: https://doi.org/10.1109/ACCESS.2022.3140421.

Oughton, E.J. and Jha, A. (2021) 'Supportive 5G Infrastructure Policies are Essential for Universal 6G: Assessment Using an Open-Source Techno-Economic Simulation Model Utilizing Remote Sensing', *IEEE Access*, 9, pp. 101924–101945. Available at: https://doi.org/10.1109/ACCESS.2021.3097627.

Pant, R. *et al.* (2020) 'Resilience study research for NIC–systems analysis of interdependent network vulnerabilities. Environmental Change Institute'. Oxford University, UK. https://www. nic. org. uk/wp-content/uploads ….

Ritchie, H., Roser, M. and Rosado, P. (2020) '$CO_2$ and Greenhouse Gas Emissions', *Our World in Data* [Preprint]. Available at: https://ourworldindata.org/co2-and-other-greenhouse-gas-emissions (Accessed: 9 July 2022).

Ruiz, D. *et al.* (2022) 'Life cycle inventory and carbon footprint assessment of wireless ICT networks for six demographic areas', *Resources, Conservation and Recycling*, 176, p. 105951. Available at: https://doi.org/10.1016/j.resconrec.2021.105951.

Schneir, J.R. *et al.* (2023) 'Guest Editorial: Techno-Economic Analysis of Telecommunications Systems', *IEEE Communications Magazine*, 61(2), pp. 22–23. Available at: https://doi.org/10.1109/MCOM.2023.10047852.

Siddik, M.A.B., Shehabi, A. and Marston, L. (2021) 'The environmental footprint of data centers in the United States', *Environmental Research Letters*, 16(6), p. 064017. Available at: https://doi.org/10.1088/1748-9326/abfba1.

Spagnuolo, A. *et al.* (2015) 'Monitoring and optimization of energy consumption of base transceiver stations', *Energy*, 81, pp. 286–293. Available at: https://doi.org/10.1016/j.energy.2014.12.040.

Stephens, H.M., Mack, E.A. and Mann, J. (2022) 'Broadband and entrepreneurship: An empirical assessment of the connection between broadband availability and new business activity across the United States', *Telematics and Informatics*, 74, p. 101873. Available at: https://doi.org/10.1016/j.tele.2022.101873.

Suman, S. and De, S. (2020) 'Low Complexity Dimensioning of Sustainable Solar-Enabled Systems: A Case of Base Station', *IEEE Transactions on Sustainable Computing*, 5(3), pp. 438–454. Available at: https://doi.org/10.1109/TSUSC.2019.2947642.

Tatem, A.J. (2017) 'WorldPop, open data for spatial demography', *Scientific Data*, 4(1), pp. 1–4. Available at: https://doi.org/10.1038/sdata.2017.4.

Thoung, C. *et al.* (2016) 'Future demand for infrastructure services', *The future of national infrastructure: A system-of-systems approach* [Preprint].

TowerXchange (2017) *TowerXchange CALA Dossier 2017*. London: TowerXchange.

TowerXchange (2018a) *TowerXchange Africa Dossier 2018*. London: TowerXchange.

TowerXchange (2018b) *TowerXchange Asia Dossier 2018*. London: TowerXchange.

TowerXchange (2019a) *TowerXchange Americas Dossier 2019*. London: TowerXchange.

TowerXchange (2019b) *TowerXchange Europe Dossier 2019*. London: TowerXchange.

TowerXchange (2019c) *TowerXchange MENA Dossier 2019*. London: TowerXchange.

Tweed, K. (2013) 'Why cellular towers in developing nations are making the move to solar power', *Scientific American*, 15.

United Nations (2023) *9.1 develop quality, reliable, sustainable and resilient infrastructure, including regional and trans- border infrastructure, to support economic development and human well-being, with a focus on affordable and equitable access for all – Indicators and a Monitoring Framework*. Available at: https://indicators.report/targets/9-1/ (Accessed: 20 February 2023).

Whitacre, B.E. (2021) 'COVID-19 and Rural Broadband: A Call to Action or More of the Same?', *Choices*, 36(3), pp. 1–10.

Williams, L., Sovacool, B.K. and Foxon, T.J. (2022) 'The energy use implications of 5G: Reviewing whole network operational energy, embodied energy, and indirect effects', *Renewable and Sustainable Energy Reviews*, 157, p. 112033. Available at: https://doi.org/10.1016/j.rser.2021.112033.

WorldPop (2019) *WorldPop :: Population*. Available at: https://www.worldpop.org/project/categories?id=3 (Accessed: 2 January 2020).

Zhang, X. (2021) 'Broadband and economic growth in China: An empirical study during the COVID-19 pandemic period', *Telematics and Informatics*, 58, p. 101533. Available at: https://doi.org/10.1016/j.tele.2020.101533.

Zhang, X. and Wei, C. (2022) 'The economic and environmental impacts of information and communication technology: A state-of-the-art review and prospects', *Resources, Conservation and Recycling*, 185, p. 106477. Available at: https://doi.org/10.1016/j.resconrec.2022.106477.